\documentclass[numberedappendix]{emulateapj}

\usepackage{multirow}
\usepackage{graphicx}
\usepackage{color}
\usepackage{longtable}
\shorttitle{Timing analysis of magnetar bursts}
\shortauthors{Huppenkothen et al.}

\begin{document}

\title{Quasi-Periodic Oscillations and broadband variability in short magnetar bursts}

\author{Daniela Huppenkothen\altaffilmark{1, 2}, Anna L. Watts\altaffilmark{1}, Phil Uttley\altaffilmark{1},  Alexander J. van der Horst\altaffilmark{1}, Michiel van der Klis\altaffilmark{1}, Chryssa
  Kouveliotou\altaffilmark{3}, Ersin G{\"o}{\u g}{\"u}{\c s}\altaffilmark{4}, Jonathan Granot\altaffilmark{5}, Simon Vaughan\altaffilmark{6}, Mark H. Finger\altaffilmark{7}}
  
\altaffiltext{1}{Astronomical Institute ``Anton Pannekoek'', University of
  Amsterdam, Postbus 94249, 1090 GE Amsterdam, the Netherlands}
\altaffiltext{2}{Email: D.Huppenkothen@uva.nl}
\altaffiltext{3}{Office of Science and Technology, ZP12, NASA Marshall Space
  Flight Center, Huntsville,
  AL 35812, USA}
\altaffiltext{4}{Sabanc\i~University, Orhanl\i-Tuzla, \.Istanbul  34956, Turkey}
\altaffiltext{5}{The Open University of Israel, 1 University Road, POB 808, Ra’anana 43537, Israel}
\altaffiltext{6}{X-Ray and Observational Astronomy Group, University of Leicester, Leicester, LE1 7RH, UK}
\altaffiltext{7}{Universities Space Research Association, Huntsville, AL 35805, USA}

\begin{abstract}
The discovery of quasi-periodic oscillations (QPOs) in magnetar giant flares has opened up prospects for neutron star asteroseismology.  However, with only three giant flares ever recorded, and only two with data of sufficient quality to search for QPOs, such analysis is seriously data limited. We set out a procedure for doing QPO searches in the far more numerous, short, less energetic magnetar bursts. The short, transient nature of these bursts requires the implementation of sophisticated statistical techniques to make reliable inferences. Using Bayesian statistics, we model the periodogram as a combination of red noise at low frequencies and white noise at high frequencies, which we show is a conservative approach to the problem. We use empirical models to make inferences about the potential signature of periodic and quasi-periodic oscillations at these frequencies. We compare our method with previously used techniques and find that although it is on the whole more conservative, it is also more reliable in ruling out false positives.  We illustrate our Bayesian method by applying it to a sample of 27 bursts from the magnetar SGR J0501+4516 observed by the Fermi Gamma-ray Burst Monitor, and we find no evidence for the presence of QPOs in any of the bursts in the unbinned spectra, but do find a candidate detection in the binned spectra of one burst. However, whether this signal is due to a genuine quasi-periodic process, or can be attributed to unmodeled effects in the noise is at this point a matter of interpretation.   
\end{abstract} 


\section{Introduction}
Neutron stars present the best test cases for extreme physics in the high-density regime.  A long-standing problem in neutron star physics is our lack of understanding of the neutron star interior, in particular, the dense matter equation of state \citep{Lattimer07}. The conditions in both a neutron star's core and crust, the composition of its matter and the topology and strength of the magnetic fields remain mysteries that are very difficult to tackle with most conventional methods.  The detection of quasi-periodic oscillations (QPOs) in the tails of giant flares from Soft Gamma Repeaters (SGRs) has opened up the possibility of studying neutron star interiors using asteroseismology (see \citealt{Watts11} for a review).  

SGRs exhibit regular bursts in the hard X-rays and soft $\gamma$-rays ($\lesssim 100 \, \mathrm{keV}$), and very rare giant flares with extremely high isotropic equivalent radiated energy of up to $10^{46} \, \mathrm{erg}$ (see e.g. \citealt{2005Natur.434.1107P}). Observations of persistent soft X-ray counterparts showing coherent pulsations with large periods of $5-8$ seconds \citep{1998Natur.393..235K, 1999ApJ...510L.115K}, and the detection of the same periodicities in the tails of the giant flares \citep{1999Natur.397...41H, 2005Natur.434.1107P}, suggested that SGRs are neutron stars.  Their behavior is understood within the context of the magnetar model \citep{1995MNRAS.275..255T}:  in this paradigm the SGRs are isolated neutron stars with exceptionally strong external dipole magnetic fields, largely above the quantum critical limit $B_c= 2 \pi m^{2}_{e}c^3/he=4.4 \times10^{13} \, \mathrm{G}$  (where $m_{e}$ is the mass of the electron, $c$ the speed of light, $h$ Planck's constant and $e$ the absolute value of the electron charge), with internal fields that may be as high as\footnote{Supported by period and period derivative measurements; see $\mathrm{http://www.physics.mcgill.ca/}$\textasciitilde$\mathrm{pulsar/magnetar/main.html}$ for an up-to-date reference list on magnetar spin-down properties} $10^{16} \, \mathrm{G}$. Giant flares are powered by a catastrophic reordering of the magnetic field \citep{2001ApJ...552..748W}.  Since this field is coupled to the solid crust, \citet{1998ApJ...498L..45D} suggested that such large-scale reconfiguration might rupture the crust, creating global seismic vibrations that would be visible as periodic modulations of the X-ray and $\gamma$-ray flux. This idea was confirmed by the detection of QPOs in the expected range of frequencies  ($\sim 10-1000$ Hz) in the tails of giant flares from two different magnetars \citep{2005ApJ...628L..53I, 2005ApJ...632L.111S, 2006ApJ...653..593S, 2006ApJ...637L.117W}. 

If the QPO frequencies can be reliably identified with particular global seismic modes of the neutron star, then they can in principle be used to constrain both the equation of state and the interior magnetic field.     This exciting possibility has driven a major effort to develop theoretical models of the vibrations.   One major issue is the effect of the strong magnetic field, which threads the crust and the core, giving rise to a spectrum of magneto-elastic oscillation frequencies that includes both continua (which give rise to unusual dynamical responses, \citealt{Levin07}) and discrete modes.  At present there is some disagreement about the nature and effects of the continua on the resulting frequencies and their longevity (see for example \citealt{vanHoven11}, \citealt{Gabler11}, and \citealt{Colaiuda12}). Uncertainties in the composition of the neutron star crust, and the role of superfluidity, will also have an effect \citep{Watts07, vanHoven08, Steiner09, Andersson09}.   How stellar vibrations cause high amplitude variations in X-ray emission is also not clear, and processes in the magnetosphere may play an active role \citep{Timokhin08, Dangelo12}.  

A major obstruction to this field of research is the sparsity of data. Since the launch of the first X-ray and $\gamma$-ray instruments, only three giant flares have been observed, with just two having data with a sufficient time resolution to detect QPOs. In trying to overcome this lack of observational constraints, it is therefore a reasonable approach to turn to the much more numerous short SGR bursts with lengths of usually less than a second and luminosities around $10^{40} \, \mathrm{erg}\, \mathrm{s}^{-1}$. Hundreds of SGR bursts have now been observed from many magnetars\footnote{see e.g. \citealt{woods06}, \citealt{mereghetti08} for overviews or http://f64.nsstc.nasa.gov/gbm/science/magnetars/ for a collection of SGR bursts observed with the Fermi Gamma-Ray Burst Monitor (Fermi/GBM)}. Additionally, several intermediate flares have been detected, with observational properties somewhere between those of the SGR bursts and those shown by giant flares (\citealt{ibrahim01}, \citealt{lenters03}, \citealt{guidorzi04}, \citealt{israelif08}). At present the nature of the trigger mechanism for both the giant flares and the shorter bursts is an open question \citep{1995MNRAS.275..255T, Lyutikov03, Duncan04, Woods05, Gill10, Perna11, Watts11, Levin12}, but it is certainly possible that the oscillations detected in giant flares may be excited in the smaller events as well. The detection of periodic signals in SGR bursts is however restricted by their length: giant flares can last up to hundreds of seconds, whereas a typical SGR burst is shorter than one second, restricting the minimum frequency that can be searched.

To date there has been no systematic search for periodic features in the lightcurves of the SGR bursts.  A search for QPOs in a period of enhanced emission with multiple bursts (a `burst storm'), from the magnetar SGR J1550-5418, carried out using data from the Fermi Gamma-ray Burst Monitor (GBM), found no significant signals \citep{Kaneko10}. \citet{2010ApJ...721L.121E} searched a subset of Rossi X-ray Timing Explorer data from SGR 1806-20 for periodic features and found some tentative signals:  however there are several points of concern with regard to their methodology which we address conceptually in Section \ref{sec:mc} and in detail in Appendix A.   

Searching for QPOs in transient light curves is a non-trivial task. Standard methods involving Fourier analysis are defined for infinitely long, stationary processes, owing to the periodic nature of the sine functions used in the Fourier transform. The very nature of a transient event - it has a start, one or more peaks, and an end - complicates the analysis procedure and introduces additional sources of uncertainty. For transient events where the shape of the burst envelope is known, many problems arising from the non-stationarity can be solved either analytically \citep{2011MNRAS.415.3561G} or via Monte Carlo simulations \citep{2001MNRAS.321..776F}. However, many astrophysical transients do not show a well-behaved burst light curve that is easily reproducible by a simple function. Magnetar bursts in particular exhibit a variety of shapes in the burst envelope, translating into different power spectral shapes in the Fourier domain, which need to be taken into account when deriving significances from the periodogram. This in itself can be interesting, aside from searching for QPOs: the different burst envelope shapes must be created by a physical process in the source, either in the form of noise processes or non-stochastic emission processes, and characterising the differences may tell us more about the emission processes at work.

This paper presents the application of a Bayesian method, first derived for long-duration time series data of Active Galactic Nuclei (AGN) in \citet{vaughan2010}, to timing analysis of magnetar bursts. We choose this method for its capabilities in finding periodicities and QPOs in red-noise dominated periodograms. However, we attempt to answer not only the question of whether there are indeed QPOs, but also to characterize the aperiodic timing behaviour of the bursts.   Given the uncertainty that exists over the trigger and emission mechanisms for magnetar bursts, such an additional diagnostic will be useful.   The method that we develop is general, in the sense that it may be applied to other transients of similar light curve morphology such as gamma-ray bursts (GRBs).

In this paper we illustrate the power of this new method by applying it to timing analysis of  a sample of SGR bursts recorded by Fermi GBM. Specifically, we search observations taken during an intense flaring episode of the SGR J0501+4516 for periodic and quasi-periodic signals, and characterise the broadband noise processes seen in the bursts. This SGR was discovered on 2008 August 22, when a burst triggered the Swift Burst Alert Monitor \citep{2008ATel.1676....1B, 2008GCN..8112....1H}. The same burst subsequently triggered the Fermi/GBM, which then recorded high time-resolution data of a total of 29 bursts over 13 days \citep{2011ApJ...739...87L}.  An RXTE Target of Opportunity pointing revealed a spin period of $~ 5.76 \, \mathrm{s}$ \citep{2008ATel.1677....1G}. With a spin-down period of $(1.5 \pm 0.5) \times 10^{-11} \, \mathrm{s} \, \mathrm{s}^{-1}$, the dipole magnetic field was estimated as $2.0 \times 10^{14} \, \mathrm{G}$ \citep{2008ATel.1691....1W, 2009MNRAS.396.2419R, 2010ApJ...722..899G}. 

In Section \ref{sec:methods}, we give an overview of the general Bayesian method of searching for periodicities and QPOs in burst data, including a comparison to previous methods. Section \ref{sec:datareduction} presents details of the instrument and the data reduction process for this burst sample. We then characterize both the method's power and limitations by applying the method first to a large number of simulated observations with and without artificially introduced periodic signals in Section \ref{sec:burstenvelope}. Subsequently, we apply the method to the Fermi GBM burst sample from SGR 0501+5416. In Section \ref{sec:results} we first outline the method using one particular burst as an example, before giving results for the whole sample. In Section \ref{sec:discussion}, we discuss the significance of our results, and put them in context with current theoretical models.
The purpose of this paper is to lay out the method and test it thoroughly on simulated data before applying it to a small burst sample to demonstrate its power on real data. In future work, it will be applied to a larger sample of short as well as intermediate bursts and giant flares.

\begin{figure*}[h!]
\begin{center}
\includegraphics[width=17cm]{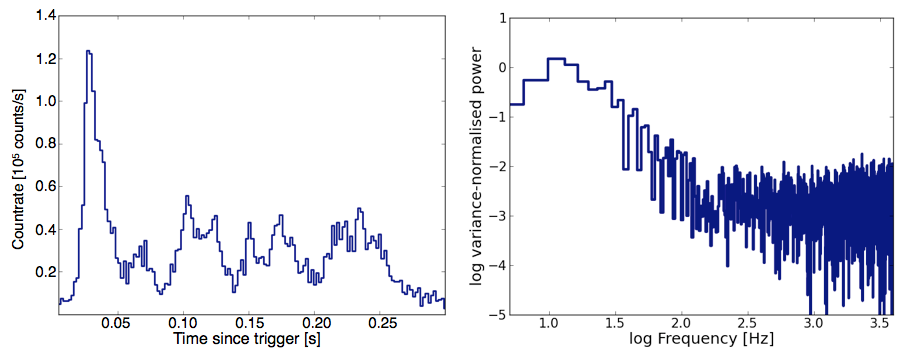}
\caption{Fermi GBM observation of burst {\it 080823478}\footnote{Fermi/GBM bursts are numbered by date in the format YYMMDDFFF with YY, the year minus 2000; MM, the two-digit month; DD the two-digit day of the month and FFF the fraction of the day} from SGR J$0501+4516$; left: light curve with a time resolution of 0.001 seconds. Structure in the burst profile and tail is clearly visible. Right: Periodogram of the burst light curve shows flat Poisson noise at high frequencies, and an excess of power over the Poisson level at low frequencies, owing to the complex shape of the light curve.}
\label{fig:bmexmp_comb}
\end{center}
\end{figure*}

\section{Variability Analysis Methods}
\label{sec:methods}

Our goals are to search for periodic and quasi-periodic signals in light curves of SGR bursts as well as to characterize the broadband variability behaviour of the bursts.
To this end, we employ Fourier techniques \citep{1989tns..conf...27V}, extending them for the special case of transient light curves and the presence of broadband variability in our burst light curves. 
Note here that, following \citet{vaughan2010}, we use the expression {\it periodogram} for the squared Fourier transform of the data. We assume that it is the sampling of the burst envelope as well as one or several noise processess.  We use the expression {\it power spectrum} to denote the underlying physical process of which the periodogram is a sample, i.e. a realization.

\subsection{Monte Carlo Simulations of Light Curves: Advantages and Shortcomings}
\label{sec:mc}

Monte Carlo simulations of light curves are a standard tool in timing analysis (see for example \citealt{2001MNRAS.321..776F}). The underlying idea is simple: one fits an empirically derived (or physically motivated) function to the burst profile. One then generates a large number of realizations of that burst profile, including appropriate sources of (usually white) noise, such as Poisson photon counting noise. The periodograms computed from these fake light curves form a basis against which to compare the periodogram of the real data. For each frequency bin, a distribution of powers is produced, with a mean that depends both on the Fourier-transformed burst envelope shape and the noise processes introduced into the light curve, while the scatter around that mean follows the noise processes only (a $\chi^2$ distribution with 2 degrees of freedom - denoted $\chi^2_2$ -  for a wide range of noise processes, as long as the central limit theorem holds). 

Comparing the observed power in each bin with the distribution of simulated powers in the same bin allows us to make a statement about the probability of the observed power in a particular bin being due to a noise process: if the observed power in a particular bin is a high outlier compared with the distribution of simulated powers in that bin, then the probability of observing the data under the (null) hypothesis of a noise process is $1/N$, where $N$ is the number of simulations performed. If N is large, the observed outlier is unlikely to be produced by the noise process alone. 

It should be noted, however, that this test only rejects the null hypothesis, it does not directly give evidence for the alternative hypothesis, i.e. the hypothesis we test the null hypothesis against, to be true. As we will explain in more detail in this section, a faulty assumption for the noise model may well produce significant detections which are, in fact, due to a noise process we have not taken into account appropriately. Conversely, a power that does not exceed the maximum simulated power may still be a significant signal, if the maximum simulated power is a rare event.

Note that the probability derived from the Monte Carlo simulations must be subjected to a correction for the number of frequencies and bursts searched (the number of trials, also called {\it Bonferroni} correction or ``look-elsewhere effect''), since for a large number of frequencies and light curves searched, we would expect a number of outliers that would otherwise be counted as (spurious) detections.

\begin{figure*}[htbp]
\begin{center}
\includegraphics[width=18cm]{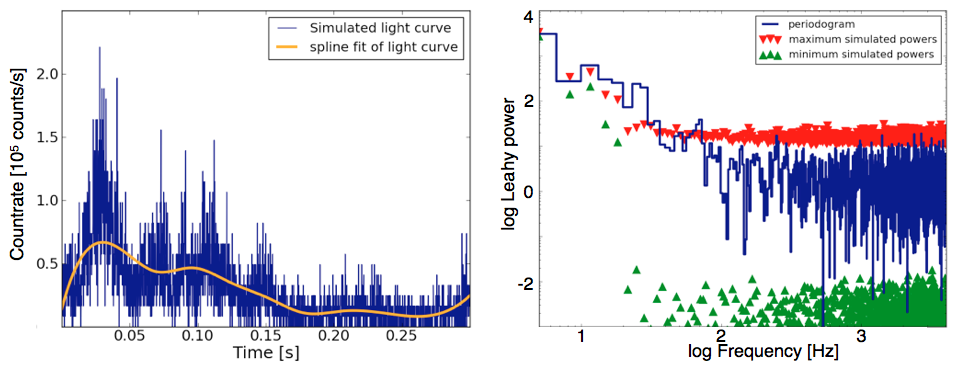}
\caption{This figure illustrates the effect that inadequate fitting of a light curve containing red noise can have on estimating the significance of potential QPO signals from Monte Carlo simulations.  Here, we followed the routine laid out in \citet{1995A&A...300..707T} to create a light curve from a red noise power spectrum with power law index of $\approx -2$ with Poisson noise added (left panel, solid dark blue line). The input model contains no periodic or quasi-periodic signal. We then binned the resulting light curve to a very coarse time resolution (ten data points) and fit the resulting binned function with a spline (left figure, orange curve). Note that the choice of resolution for the fit is arbitrary: we cannot know a priori from the light curve which features are created by a broadband noise process or a quasi-periodic signal. The binned light curve was used as the basis for Monte Carlo simulations. The figure on the right side shows the periodogram of the fake light curve (dark blue line), with the maximum (red downward triangles) and minimum (green upward triangles) simulated power in each bin. The maximum and minimum powers at each frequency were found by sampling the distribution of powers at that frequency in 1000 Monte Carlo simulations of the light curve fit (orange line in the left panel) with added Poisson noise, i.e. not taking into account red noise, and taking the minimum and maximum samples. Note the spurious detections at 25 Hz and 70 Hz, where peaks in the periodogram clearly stick above the distribution of expected noise powers, even though there is no QPO feature at this frequency: it is entirely created by red noise.}
\label{fig:mc1}
\end{center}
\end{figure*}

The Monte Carlo method outlined above is versatile and powerful, but it has limitations. The most important limitation comes from our lack of knowledge of the underpinning physical processes producing the observed light curve. Only if the null hypothesis accurately reflects the data - apart from the (quasi-)periodic signal for which we would like to test - is the test meaningful. If important effects that distort either shape or distribution of the powers are missed, then the predictions made will not be accurate, leading to either spurious detections or real signals not being found.

More often than not, especially in the case of short magnetar bursts, we do not have complete information about the emission mechanism. Short magnetar bursts are extremely diverse, varying in light curve shape as well as burst intensity and duration (see, for example, \citealt{1999ApJ...526L..93G} and  \citealt{2000ApJ...532L.121G}). Unlike for thermonuclear X-ray bursts, where the Monte Carlo technique is widely employed (see for example \citealt{2001MNRAS.321..776F} and \citealt{Watts05}), we do not know the underlying aperiodic shape of the light curve (see e.g. Figure \ref{fig:bmexmp_comb} for an illustration of a complicated SGR light curve and periodogram). In order to apply this method, we therefore have to fit the light curve using a parametric approach involving e.g. higher-order polynomials or splines, and then use this as a template for the Monte Carlo simulations. There is an essential degeneracy in that problem originating from our lack of knowledge: which features do we fit? Which do we consider to be part of the burst envelope, or potential candidates for a periodic signal on top? This is an arbitrary decision, however one that greatly influences the probability of detecting spurious signals. 

The situation is further complicated by the potential presence of so-called red noise: random processes that produce broadband, aperiodic variability, frequently with power-law type shapes in the Fourier domain. Red noise supplies large powers at low frequencies and little at high frequencies, and a realization of a red noise process can appear to the naked eye in the light curve like a possibly periodic process (the four peaks in the light curve of Figure \ref{fig:bmexmp_comb}, for example, seem to mimic periodic behaviour, but there is only a very broad bump in the periodogram, and it is impossible to distinguish between a broken power law and a very broad quasi-periodicity in this case). The presence of red noise will significantly alter the distribution of powers in each frequency bin from what it would be if the light curve consisted of a purely deterministic envelope and Poisson-distributed detector noise.
Thus, even when the shape of the burst envelope can be adequately modeled by a single, deterministic function, there may nevertheless be false positive detection of single-bin QPO features which are purely due to scatter in low-frequency bins owing to the presence of red noise that has not been taken into account.

In Figure \ref{fig:mc1}, we set up a model that contains only red noise, and find a significant detection despite the fact that there was no QPO introduced into the light curve.
This illustrates the fundamental problem with simulating light curves that have an unknown underlying shape. Features may be due to broadband noise features in the light curve, which are not accurately represented by the coarse light curve, and thus not adequately modeled by our null hypothesis. Hence, these features are flagged as a significant detection (with a single-trial probability of $10^{-3}$), despite not being due to an underlying (quasi-) periodic process. In the absence of a well-known underlying burst envelope shape or physically motivated models of both noise and burst envelope, it is thus not advisable to apply this method to magnetar bursts (or any transient events with complex light curve shape) in order to derive meaningful conclusions about the presence of QPOs in the light curve. This is one of several shortcomings of the procedure presented in \citet{2010ApJ...721L.121E}. We comment on this paper in the context of our new procedure in greater detail in Appendix A.  \\
We note that the distinction between QPOs and some forms of noise is not clear cut. By convention, one usually defines an upper boundary for the full-width half-maximum FWHM $ < \nu_0/2$ (where $\nu_0$ is the QPO's centroid frequency, \citealt{2006csxs.book...39V}) for the feature to be called a QPO, however, this is a somewhat arbitrary decision. In this work, we consider aperiodic noise only in the form of power laws or broken power laws, as opposed to QPO features which we assume to be fairly narrow features (following the convention for the FWHM mentioned above) on top of this process. It should be noticed that in principle, one could fit a broadband feature with a wide Lorentzian, thus there is some degeneracy in the modeling.  
In Section \ref{sec:bayes}, we explore whether other simple, plausible models can fit the data, and will describe an alternative, conservative method, based on the assumption that red noise dominates the power spectrum. This is unlikely to be completely true, although many bursts seem to have a red noise component of varying strength, but as we will lay out in the following sections, this assumption is less prone to producing false positive detections, at the cost of increased risk of false non-detections.

\subsection{Modeling the Periodogram}
\label{sec:bayes}
Another approach to the problem uses models of the observed periodogram rather than the light curve and assumes any low-frequency broadband variability to be due to a noise process. In a way, this is the other extreme to the approach of using Monte Carlo simulations of light curves: the former is based on the null hypothesis that any power in the periodogram is due to a combination of a deterministic burst envelope, photon counting (white) noise and a putative (quasi-) periodic signal that is to be detected. When modeling the periodogram, we instead assume that there is no deterministic contribution from the burst envelope and the entire observed emission is due to a noise process. Unless the emission process itself is a noise process, this may not be a valid assumption either. In effect, assuming pure red noise while the light curve has both a noise component and a non-stochastic envelope will cause us to miss weak signals at low frequencies, because they are buried in the much higher variance at low frequencies in a broadband noise process compared to a deterministic burst envelope with only white noise on top.

For the power spectral modeling, we closely follow the Bayesian approach developed by \citet{vaughan2010}. One advantage of the Bayesian framework is the inclusion of our uncertainty in the model parameters (of the assumed low-frequency noise process) in the error estimate of any final quantity, although this still assumes that the functional shape of the spectrum is known; this must be determined separately, as we will lay out below. In addition it provides a statistically rigorous framework to test whether additional model components (such as Lorentzian QPOs) are required by the data \citep{2002ApJ...571..545P}. In the following, we give only a short outline of the method, and refer the reader to \citet{vaughan2010} for a thorough discussion. 

Following Bayes' rule, the posterior probability of a set of model parameters $\mathbf{\theta}$ of interest, given the observed data $\mathbf{I}$  and a model $H$, is defined as

\begin{equation}
p( \mathbf{\theta} | \mathbf{I}, H) = \frac{p(\mathbf{I} | \mathbf{\theta}, H) p(\mathbf{\theta}| H) }{p(\mathbf{I} | H)} \, .
\label{eqn:bayesrule}
\end{equation}

Here, $p(\mathbf{I} | \mathbf{\theta}, H)$ is called the {\it likelihood}, $p(\mathbf{\theta}| H)$ the {\it prior} and $p(\mathbf{I} | H)$ the {\it marginal distribution} of the data. Note that the latter is often difficult to compute in practice, and only depends on the data. For ratios of posterior probabilities utilising the same data, the marginal distribution of the data will drop out of the equation, and will consequently be ignored in the following. 

We use the Bayesian analogue to maximum likelihood estimation (MLE) to fit models to the observed periodogram data and obtain the Maximum A Posteriori (MAP) estimates of the model parameters. 
The MAP estimate of a parameter set is defined as:

\begin{equation}
\mathbf{\theta}_{MAP} = \arg\max_{\theta} p( \mathbf{\theta} | \mathbf{I}, H) \; ,
\label{eqn:map}
\end{equation}

where $\arg$ is the argument of the maximized posterior probability, i.e. $\theta_{\mathrm{max}}$.
The MLE of a given model $S(\mathbf{\theta})$ is computed by maximizing the probability of a data set $\mathbf{I}$ given parameters $\mathbf{\theta}$ and a model $H$:

\begin{equation}
p(\mathbf{I} | \mathbf{\theta}, H) = \prod_{j=1}^{N/2}p(I_j | S_j) \, ,
\label{eqn:likelihood}
\end{equation}
where $I_j$ are the individual powers in the observed power spectrum, and $S_j$ are the powers in the chosen models for a parameter set $\mathbf{\theta}$.
This is equivalent to minimizing the following function:

\begin{equation}
D(\mathbf{I}, \mathbf{\theta}, H) = - 2 \log{p(\mathbf{I} | \mathbf{\theta}, H)} = 2 \sum_{j=1}^{N/2} \left\{ \frac{I_j}{S_j} + \log{S_j}  \right\} \; ,
\label{eqn:deviance}
\end{equation}
sometimes called the deviance \citep{gelman2004}. Note that Equation \ref{eqn:deviance} is only a valid form of Equation \ref{eqn:likelihood} if the data are $\chi^2$-distributed. 

Similarly, we can compute the logarithmic MAP as a combination of prior and likelihood, using Equations \ref{eqn:bayesrule}, \ref{eqn:map} and \ref{eqn:deviance}:

\begin{eqnarray}
\mathbf{\theta}_{MAP} & = &  \arg\max_{\theta}( p(\mathbf{I} | \mathbf{\theta}, H) p(\mathbf{\theta}| H))  \\ 
 & =  & \arg\min_{\theta} (- \log p(\mathbf{\theta}| H)   + D(\mathbf{I}, \mathbf{\theta}, H)/2)   \; . \label{eqn:mapwdeviance}
\end{eqnarray}

Equation \ref{eqn:mapwdeviance} computes the mode of the posterior distribution over parameter space, i.e. the most likely parameters given the observed data and the model. We use the formalism above for any analysis done in this work. 

Without a physically motivated burst emission mechanism, we cannot know what shape the analytic part of the burst envelope takes, or the existence and characteristics of a potential red noise component in the data. Since both the burst envelope and any red noise processes supply power over a large range of frequencies (unlike QPOs, which are confined to narrow regions of frequency space), there is an essential degeneracy in any model we attempt to fit, adding a number of assumptions about the burst shape and red noise properties to whatever inference we try to make. In the absence of any physical motivation, we make a simple, yet probably overly conservative assumption: all broadband power in the periodogram is supplied by a red noise process. The limitations on our inferences that come from this assumption will be further discussed in Section \ref{sec:burstenvelope}, but its main disadvantage is the fact that weak signals in the low-frequency part of the spectrum are likely to be missed. We see this as an acceptable trade-off in return for having a very low false positive detection rate. 
The advantage of this assumption is that we can treat the entire broadband variability seen in the periodogram as a realization of a noise process and follow an entirely empirical approach to modeling the periodogram: if we find a function that describes the underlying power spectrum well, we can use this as a basis to compute many realizations of this power spectrum and compare these to our observed data.
A very broad class of power spectral shapes well represented in nature are power laws:

\begin{equation}
P(\nu) = \beta \nu^{-\alpha} + \gamma \; ,
\label{eqn:pl}
\end{equation}
where $\alpha$ is the power law index, and broken power laws, which can be reduced to Equation \ref{eqn:pl} by setting $\rho = 0$:

\begin{equation}
P(\nu) = \beta \nu^{-\alpha_1} \left( 1 + \left\{ \frac{\nu}{\delta} \right\}^{(\alpha_2 -\alpha_1)/\rho}  \right)^{-\rho} + \gamma \; ,
\end{equation}
where $\alpha_1$ and $\alpha_2$ are the power law indices at low and high frequencies, respectively, and we require $\alpha_2 < \alpha_1$. $\delta$ is the break frequency at which the power law index changes. In both models, $\beta$ is a normalization term and $\gamma$ is a constant to account for the presence of white (Poisson) noise in the periodogram. Note that the broken power law is a more general expression of the bent power law used in \citet{vaughan2010}, including the additional smoothness parameter $\rho$ to account for the smoothness of transition between the two power law components. 

Our lack of knowledge of the emission processes in magnetar bursts leads us to choose uninformative prior probability distributions for all model parameters: a $p(\theta) = 1/\theta$ dependence for all scale parameters $\beta$, $\gamma$ and $\delta$ (a Jeffreys prior, see \citealt{vaughan2010} and references therein), and flat priors $p(\theta) = \mathrm{constant}$ for all other parameters. 
Together, these two classes describe a large range of broadband variability, and are likely to be sufficient in describing the low-frequency behaviour of most magnetar burst periodograms. In what follows, we choose our broadband noise model from one of these two.  However we include an overall goodness-of-fit test and comment where the model fails this test for individual bursts. 

There are several ways to distinguish between models. One often used statistic for nested models, i.e. models where one is a special case of the other, is the Likelihood Ratio Test (LRT).  The LRT statistic is based on the ratio of the likelihood values for the two models, the {\it null hypothesis} $H_0$ and the {\it alternative hypothesis} $H_1$:

\begin{eqnarray}
T_{LRT }  & = & -2 \log{\frac{p(\mathbf{I} |\hat{\mathbf{\theta}}^0_{MAP} , H_0 )}{p(\mathbf{I} |¬†\hat{\mathbf{\theta}}^1_{MAP} , H_1 )}}  \\ \nonumber
 		& = & D_{min}(H_0) - D_{min}(H_1) \; .
\end{eqnarray}

In order to decide whether a data set is adequately described by the null hypothesis or not, one often resorts to Monte Carlo simulations of the null hypothesis.
In the Bayesian framework, one can employ a certain type of Monte Carlo simulations, so called Markov Chain Monte Carlo simulations (MCMCs), to draw parameter sets from the posterior distribution of possible parameters and generate predictive (fake) periodogram data this way. MCMCs have the advantage that for a stable chain that has converged, the samples generated in that chain will always approximate the posterior distribution of parameters,  i.e. the distribution for each parameter that summarizes our entire knowledge of the problem. The posterior distribution for each parameter is obtained by marginalizing (i.e. integrating) over all other model components. In the case where the parameter distributions are non-Gaussian, this allows for far more accurate modes and errors on the individual parameters than standard methods like the covariance matrix. Probably the most widely known and employed MCMC algorithm is the Metropolis-Hastings algorithm \citep{metropolis:1087, HASTINGS01041970}. However, in many situations, convergence of this algorithm is slow and hence computationally expensive. In this work, we employ the so-called stretch-move algorithm as implemented in python by \citet{2012arXiv1202.3665F} in the module {\it emcee}. {\it emcee} uses so-called ensemble walkers: a set of Markov chains that is split in two, where each half is evolved using the state of the other half as an input, thereby increasing efficiency in converging towards the posterior distribution of parameters. 

The MCMC produces a sample of parameter values (of the null hypothesis, e.g. a continuum model) drawn from the posterior distribution of the data. From this sample we randomly draw parameter vectors and use these to generate fake periodograms. We can then compute a distribution for a statistic $T$ to compare with the same statistic derived from the observed data, $T_{\mathrm{obs}}$. In the case of model selection, the faked data are fit with both a simple and a more complex model (e.g. a power law and a broken power law), $H_0$ and $H_1$, identical to the procedure performed on the observed data. This generates a distribution of $T_{LRT}$s, which is then used to calculate the corresponding tail area probability (i.e. the probability of obtaining a value of the test statistic that is at least as extreme as the one observed under the assumption of the null hypothesis, also called p-value) for the observed value of $T^{obs}_{LRT}$. If this probability is very small (the actual detection level is subject to choice, for example $p < 0.05$), then the observed reduction in $D_{min}$ between $H_0$ and $H_1$ is larger than can be expected by chance if $H_0$ were true. More clearly, we reject the null hypothesis in this case, although this test cannot be seen as direct evidence that the alternative hypothesis is true. Conversely, if the probability is not very small, then $H_0$ is sufficient to describe the data. 

Just as data were simulated for assessing the probability of $T_{LRT}^{obs}$, we can generate fake data in the form of MCMCs to calculate the distribution of any test statistic we choose.
One is particularly sensitive to the kind of model features we are interested in detecting, namely breaks/bends in the smooth continuum, in that it indicates whether the model provides a good overall fit to the data, or whether additional model components may be needed.
This simple statistic for goodness-of-fit of aperiodic features, based on the traditional $\chi^2$ statistic, i.e. the sum of the squared standard errors \citep{1990ApJ...364..699A, vaughan2010}, is

\begin{equation}
T_{SSE} = \chi^2(\mathbf{I}, \hat{\mathbf{\theta}}) \; ,
\end{equation}
where

\[
\chi^2(\mathbf{I}, \mathbf{\theta}) = \sum_{j=1}^{N/2}{\frac{(I_j - E[I_j | \mathbf{\theta}] )^2}{E[I_j | \mathbf{\theta}]}} = \sum_{j=1}^{N/2}{\left( \frac{I_j - S_j(\mathbf{\theta})}{S_j(\mathbf{\theta})} \right)^2} \; 
\]
and $E[]$ indicates expectation. This is a good test of overall goodness-of-fit which will be sensitive to inadequacies in the continuum modeling since all data points are included (as opposed to the $T_R$ statistic, i.e. the biggest outlier in the data, which we will present below in Section \ref{sec:qpos}).  \\

We have now characterized the broadband noise properties. This information will be the basis for any modeling of the data done in the remainder of this section. In the following, we define a test statistic for outliers in the data, show how to compute posterior predictive p-values for this statistic, and lay out a method to find broader signals, i.e. QPOs, in the data.

\subsubsection{Searching for (Quasi-)Periodicities}
\label{sec:qpos}
The procedure for searching for periodicities and QPOs in the data follows the same basic logic applied to the selection of a broadband noise model above.  We compute a statistic from the periodogram, then generate a large number (e.g. $1000$) of simulated periodograms from an MCMC sample, compute the desired statistic from each simulated periodogram in turn and finally compare the observed value of the statistic to the distribution generated from the sample of simulations. 

In what follows, we have to distinguish very narrow features (with scale parameter, i.e. half-width half maximum (HWHM) smaller than or close to the frequency resolution of the periodogram) from broader QPO signals with HWHM that are significantly larger than the periodogram's frequency resolution. 

In order to investigate narrow features, a sensible statistic to use is the maximum ratio of observed to model power, or

\begin{equation}
\label{eqn:trstat}
T_R = \mathrm{max}_j(\hat{R_j}) \; ,
\end{equation}
where 

\[
\hat{R}_j = 2 I_j/S_j  \,
\]
and $I_j$ and $S_j$ are observed and model powers as defined above. The factor of $2$ normalises the residuals in such a way that $\hat{R}_j $ will be distributed as $\chi^2_2$. Drawing from many MCMC simulations, we can compute the tail area probability of $T_R$ from its distribution, or the probability that the observed power $I_{j, max}$ was produced purely by noise generated by a broadband model.  This probability need not be corrected for the number of frequencies searched, as this is already taken into account by the fact that we search the entire frequency range for each simulated periodogram, but if several bursts are searched, it is necessary to correct for the number of bursts searched. 

Using the posterior distribution of $T_R$ from the simulations, we can also easily compute the sensitivity to a periodic signal that could have been present in the data, but would have been missed. Sensitivities will be independent of frequency in the white noise range, but strongly depend on frequency in the red noise range, for a simple reason: a signal that would be highly significant in the white noise range could be buried under strong red noise of equal or larger strength in the low-frequency part of the spectrum, rendering it invisible to our detection methods.
We compute sensitivities for the amplitude of a potentially missed periodic signal by finding that value of $T_R$ in our simulated posterior distribution which corresponds to a posterior predictive p-value of 5\% or lower. We then compute the corresponding signal powers $I_j = R_j S_j/2$ and convert these to fractional rms amplitudes at four representative frequencies - 40 Hz, 70 Hz, 100 Hz and 500 Hz -, two of which are, for typical magnetar bursts, in the red noise dominated part of the spectrum, one right on the boundary to white noise and the last safely in the white noise dominated part of the spectrum. It is, in principle, possible to compute sensitivities for every frequency in the periodogram, however for brevity we decided to restrict ourselves to four frequencies where QPOs may be found as an indicator of the rms amplitude a signal would have to have in the different parts of the spectrum in order to be detectable. 
Note that the sensitivities computed here are different from an upper limit in the sense that they do not require the actual observation of the highest power in the spectrum: the quantity is derived entirely from the simulations, and thus presents a theoretical upper limit to what we could have measured, independent of what we have actually measured in the observed burst itself (see \citealt{kashyap2010} for a discussion on the real meaning of upper limits). \\

One shortcoming of the $T_R$ statistic is that it optimally detects periodic signals confined to one frequency bin, i.e. either strictly sinusoidal signals or QPOs with a width that is smaller than the frequency resolution of the periodogram. It should be noted that even a strictly sinosoidal signal will distribute power in more than one bin, unless its frequency is exactly the Fourier frequency. This redistribution of the power can account for an average loss of roughly $30\%$ (assuming random distribution of the sinusoidal frequency within a bin) in the frequency bin containing the sinusoid. 
Broader signals may well be detected, if they are strong enough, but since the power is spread over several bins, this is not an optimal way of detecting broad signals. There are several ways around this restriction. One is to bin (or smooth) the data in some way, and compute $T_R$ for the binned data, assuming that any tentative signal power will now be concentrated in each bin. If we bin the simulated periodograms in the same way, then the test statistic $T_R$ for the binned data is comparable to the distribution approximated by our simulations, and the latter can be used to derive posterior predictive p-values.  One can either bin the periodogram with several frequency resolutions and search for QPOs in each, assuming that for a QPO of a given width, all its power will be confined to the central one or two frequency bins if the frequency resolution is coarse enough. Alternatively, one can bin the periodogram geometrically, where the bin size grows with frequency.  This way, using the (fairly arbitrary) definition that a QPO must have a full-width half maximum $\Delta \nu$ narrower than $\nu_0/2$, with $\nu_0$ the centroid frequency of the QPO, (see e.g. \citealt{2006csxs.book...39V}), one accounts for the fact that QPOs at higher frequencies can have a larger range of widths.

An entirely different approach to the problem, which we also include in our analysis, starts out from a model selection point of view, addressing it in a similar fashion to the way one chooses between broadband noise models. Assuming that a quasi-periodicity is simply another type of random process, one may fit the periodogram simultaneously with a broadband noise process as well as a Lorentzian representing a QPO and compare the resulting fit with that of the broadband noise model only.
Following \citet{2002ApJ...571..545P}, we can utilise the likelihood ratio in this case if we compare it to the distribution of likelihood ratios as approximated by MCMC simulations. It is important to note that fitting narrow features with a Lorentzian is statistically challenging \citep{2008ApJ...688..807P, 2012ApJ...746..131B}.  For quasi-periodic features broader than a single bin, but only distributed over a few bins, we smooth the periodogram with a Wiener filter over 3, 5 and 11 frequency bins and compare the maximum power in each of the resulting smoothed spectra via the same method used for searching for single-bin periodicities presented above. Subsequently, we use the method of fitting Lorentzians only for features broader than 5 times the frequency resolution of the periodogram. Additionally, we cross-check our detection method for QPOs by searching binned spectra of lower frequency resolution than the original periodogram.

We begin by fitting a Lorentzian plus a constant to the residuals of the data divided by the preferred broadband noise model. 
At each frequency, we fix the centroid of the Lorentzian to that frequency and let the code fit the scale parameter (HWHM) and the normalization of the Lorentzian. This way, we generate an estimate of the deviance at every frequency.  
The frequency where the MAP estimate is largest we define as our tentative QPO identification. Note that we use a Jeffrey's prior ($p(x) \propto 1/x$) on the QPO normalization, and a flat prior on the QPO HWHM that rules out widths outside the range $5\Delta\nu$ to $\nu_0/2)$, where $\Delta\nu$ is the frequency resolution of the periodogram and $\nu_0$ the frequency at which we are currently fitting (the centroid frequency). This should help us avoid some of the problems with fitting narrow signals as laid out in \citet{2008ApJ...688..807P}. Restricting ourselves to HWHM larger than $5\Delta\nu$ is consistent with our choice of fitting the smoothed data: we do not expect the HWHM to be narrower than the binning we have chosen, thus we do not allow optimization to smaller widths.
Additionally, we exclude the first and last three frequency bins in order to avoid effects introduced by trying to fit a Lorentzian to one of the edges of the periodogram.

Subsequently, we combine the results from the broadband model fitting and the residual fit that yielded the highest estimate for the deviance, and use both as a starting point for a mixed model to the observed data. We use the same priors as before, but use the best-fit parameter sets of the broadband model fitting and the residual fitting as inputs to the optimization routine. We expect  that this will put us fairly close to the global minimum and help us avoid some of the problems associated with trying to minimize a multimodal likelihood function. 

Finally, we form a likelihood ratio between the broadband plus QPO model and the broadband model alone. The above procedure is repeated in exactly the same way on a large sample of MCMC generated fake periodograms in order to produce a distribution of likelihood ratios from the broadband model alone. Comparison of the observed likelihood ratio then allows the derivation of a tail area probability that the observed tentative QPO could be generated from the broadband noise model alone.
It should be noted that because the model fitting of Lorentzians on the residuals on many simulated periodograms is computationally expensive, we restrict the analysis to a smaller number of simulations (usually $N \sim 500$). The resulting distribution of likelihood ratios will be less reliable, but reliable enough to rule out most cases where there is no QPO present. If the fraction of simulations exceeding our criterion follows a binomial distribution, we can compute the error on the p-value from the standard deviation of our p-value estimate: for $p < 0.05$ and 500 simulations, the error on the p-value is $\Delta p = \sqrt{p*(1-p)/N} = 0.0097$.
For all borderline cases where the posterior predictive p-value drops below $\sim 0.1$, we repeat the analysis with a larger number of simulations ($\sim 1000$, decreasing the error on the p-value to $\Delta p = \sqrt{p*(1-p)/N} = 0.0069$) to make our estimate of the posterior p-value on the likelihood ratio more reliable. Since the error on the p-value is not high enough to bring a signal at the $5\%$ level up to $p = 0.1$, we should be able to catch all significant QPOs in this way.

\subsection{Summary of Procedure}

The Bayesian procedure laid out above has three parts: (a) find the preferred broadband noise model to represent the low-frequency part of the periodogram, (b) search the periodogram for the highest outlier and compare this outlier to those distributed by pure broadband noise to find narrow features, (c) search for QPOs in the data, using binned data as well as an identical approach for the model selection in the first step. A step-by-step description can be found in Appendix \ref{sec:recipes}.

Every step in the analysis follows the same logic: assume a null hypothesis and an alternative hypothesis, compute statistics to summarise the data-model fits for the two different models, generate a sample from this null hypothesis using MCMC, then compare the distribution of the relevant statistic derived from the sample generated from the null hypothesis to the observed value of that statistic.
If the observed value lies in the high-end tail of the distribution, then it is an outlier with respect to the null hypothesis. 

Since the entire procedure rests on the correct choice of broadband model, this is the first step of the analysis. The data are fitted with two continuum noise models, which, by definition of the likelihood ratio test, are required to be nested. The likelihood ratio is the statistic we use to decide which model is preferred by the data. We simulate a large number of fake periodograms from parameter sets drawn from the posterior distribution of parameters, as approximated by a large number of MCMC simulations. Then these fake periodograms are fit with both models again to build a distribution of likelihood ratios from the simple model. We can compute the tail-area probability (p-value) of the observed likelihood ratio to be typical of the distribution (equivalent to asking whether the observed data is sufficiently described by the simpler model) by integrating over the tail of the distribution. If this probability is lower than a chosen significance threshold, then the data is more likely to be drawn from the more complex model hypothesis, which should then be adopted for the rest of the analysis.

\begin{figure*}
\begin{center}
\includegraphics[width=14cm]{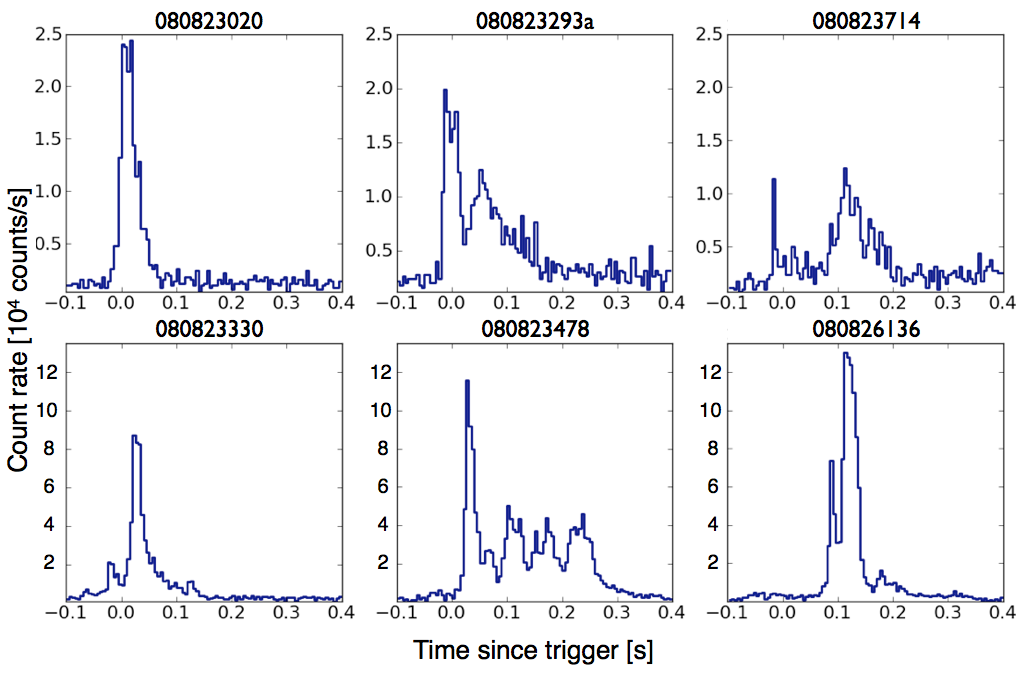}
\caption{Light curves of six example bursts from the magnetar SGR J$0501+4516$ recorded by Fermi/{\it GBM}. We combined data from all NaI detectors with source angles smaller than $50$ degrees to the source. The time resolution corresponds to $0.005$ seconds. Note the strong component of aperiodic variability after the main burst in $080823478$ and the differences in peak count rate by almost one order of magnitude between the upper three bursts and the lower three.}
\label{fig:burstselect}
\end{center}
\end{figure*} 

Finding periodicities with widths equal to or smaller than a single bin (both smoothed or unsmoothed) follows the same principle. We find the highest outlier in the residuals of the data divided by the best-fit broadband model, and compare this to a distribution of outliers computed in the same way from a large number of periodograms created from an MCMC sample. These fake periodograms do not have a periodic signal (our null hypothesis), thus if the observed outlier were far away from the simulated distribution of outliers, it is unlikely that the outlier has come from this distribution, and we hence favour the alternative hypothesis: that the outlier was indeed produced by a separate physical process.

Finally, we approach the QPO search as a model selection problem. In order not to bias ourselves to a frequency, we fit a Lorentzian at every frequency to the periodogram residuals smoothed over five bins, keeping the centroid frequency of the Lorentzian fixed while allowing the other parameters to vary. This will give us the MAP estimate for that model at each frequency. We pick the frequency with the highest MAP estimate, and fit a combined broadband model plus Lorentzian to the actual data set. In this case, however, we leave all parameters free, although we use the best-fit parameters from the residual fit as input to the simulations. This minimizes the risk of getting stuck at a local maximum of a multimodal likelihood function. This way, we can compute a likelihood ratio between the broadband noise model with an added Lorentzian component to the pure broadband model fit.
We repeat this procedure on a large number of fake periodograms without a QPO and compare the distribution of likelihood ratios to the observed likelihood ratio. Again, if that probability is very small, the observed data are unlikely under the null hypothesis, and the observed feature is unlikely the result of a chance fluctuation from an aperiodic noise spectrum alone.

\section{Data Reduction}
\label{sec:datareduction}

We now turn to a sample of magnetar bursts and illustrate our method on simulations as well as a small dataset as described below.

\subsection{Fermi/GBM}
The Gamma-ray Burst Monitor (GBM) is one of two instruments on board the Fermi Gamma-ray Space Telescope, launched in June 2008 \citep{2009ApJ...702..791M}. With its wide field of view and continuous broad-band energy coverage between 8 keV and 40 MeV, Fermi GBM is well-suited for observing magnetar bursts. The instrument triggers on magnetar bursts, providing high time-resolution data for 30 seconds before and up to 300 seconds after the trigger. Three data types were routinely output: CTIME data provide a higher time resolution (64 ms), but low energy resolution (8 channels), whereas CSPEC data provide high energy resolution (128 channels) at low time resolution (1024 ms). Note that CTIME and CSPEC data are available in lower resolution continuously; the quoted numbers are valid for trigger mode only. In this paper, only data of the third type, so-called time-tagged event (TTE) data, were used, since they provide the high time resolution ($2 \, \mu s$) required for timing analyses, while retaining full spectral resolution as well. For a detailed description of the available data modes and their properties, see \citet{2009ApJ...702..791M}.

 \subsection{Observations}
Fermi/{\it GBM} triggered 26 times on SGR J$0501+4516$ between 2008 August 22 and 2008 September 03, observing 29 bursts. Two of these ({\it 080824054} and {\it 080825200}) had saturated parts, and were therefore excluded from the analysis due to the rather complicated effects saturation can have on periodograms. Following \citet{2011ApJ...739...87L}, we used only NaI detectors with an angle to the source smaller than $50\deg$ for each of the 24 triggered and 3 untriggered bursts. The data were barycentered and channels converted to the mid-energy of each energy bin. The observations were then energy-selected to include only counts between $8$ and $100$ keV. The lower limit to the energy is set by the detector response \citep{2009ApJ...702..791M}, the upper limit was found by inspecting energy-resolved light curves and finding no source counts above 100 keV (as indicated by the counts being consistent with the Poisson distribution expected from counting noise). 
Burst start times and lengths (T90 durations) were taken from \citet{2011ApJ...739...87L}, and are summarized in Table 1 of that paper. We added $20\%$ of the burst duration to both ends of the burst in order to ensure that we caught the entire burst, and all burst start times and durations in the remainder of this article are to be understood this way.   A selection of six bursts is shown in Figure \ref{fig:burstselect}, to emphasize the diversity of burst morphologies we encounter.

\section{Detectability Simulations}
\label{sec:burstenvelope}

We test the power of our detection method on a large number of fake observations: light curves with or without a burst envelope, one or several noise processes and a periodic signal. We restrict ourselves in the following to detecting periodic or narrow quasi-periodic signals for reasons of computation time, since the QPO detection method introduced in Section \ref{sec:qpos} is computationally expensive and hence unfeasible to run on the large number (of the order of several thousand) fake light curves employed here to understand the effect of different components in the light curve on detectability of periodic signals.

While in the previous section we laid out the general principles of the method, our main goal in this section is to characterize how our assumption of pure red noise influences detection rates when this assumption is not true, e.g. in light curves with a strong burst envelope, or when the assumption holds, i.e. in light curves that contain only red noise. We start out with a simple estimate for the importance of the burst envelope on the statistical distribution of the observed powers in our burst sample, and then use one burst from our sample as a template for extensive simulations of light curves into which we artificially inject a periodic signal of varying fractional rms amplitude. 
 
 \subsection{A Simple Estimate}
 
The method laid out in Section \ref{sec:mc} is based on the assumption that an observed light curve consists of a deterministic burst envelope - a window function of some kind - and Poisson noise originating in the quantum nature of light when photons impinge on the detector. One may view the deterministic envelope as a physical process giving rise to the overall shape of the burst, following the same or at least similar functional dependencies for potentially all bursts, and, more importantly, not a realization of a noise process that would alter the general shape of the burst significantly in a stochastic way. This sets it apart from other processes we consider, which contribute to the light curve in a stochastic way. Note, however, that the characteristics given above do not imply that the burst envelope itself may be a realization of a stochastic process, with variable parameters between bursts.
The combination of burst envelope and Poisson noise is the null hypothesis against which one wishes to test. One must then ask which part of the light curve is supplied by the burst envelope, and what could be due to a potential periodic process. The presence of red noise clearly renders the fundamental assumption of this method invalid. Assuming pure red noise, on the other hand, lets us avoid a question we cannot easily answer: how much of the observed light curve can be attributed to the burst envelope, and how much to a potential noise process. We do not know {\it a priori} what the shape of the burst envelope might be, nor what the power spectral density of the noise process looks like. To first order, we already impose a window function on the periodogram simply by having a short burst: the light curve we Fourier transform is short, equalling a window function that is one between start and end times of the burst and zero everywhere else.

\begin{figure}[]
\begin{center}
\includegraphics[width=9cm]{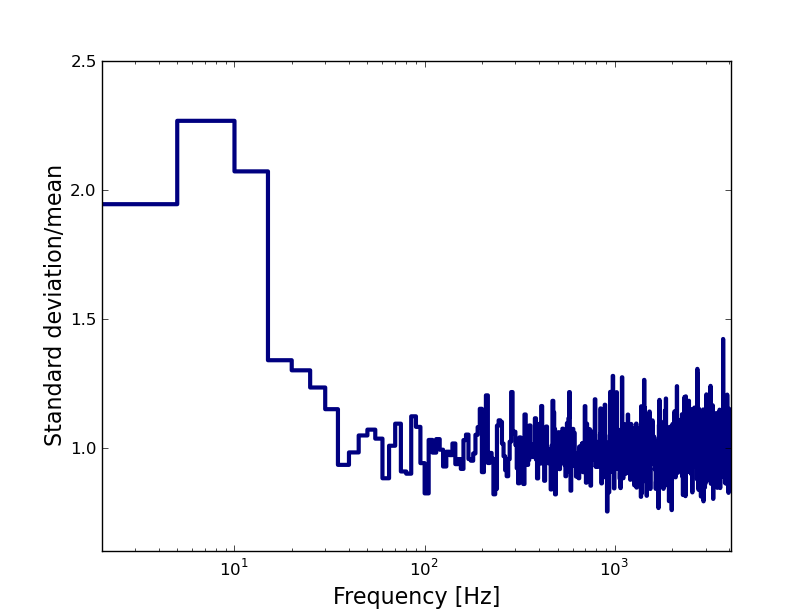}
\caption{Variation between Leahy-normalised periodograms of 27 magnetar bursts. We stretched each light curve to normalise all to the same burst duration by multiplying the photon arrival times by a scaling factor, and then computed periodograms for each of the 27 bursts in our sample. We then gathered powers in 5 Hz bins to yield distributions of powers from all bursts for each of the 5 Hz bins. For the pure red noise assumption to be valid, the resulting distributions should follow a $\chi^2_2$ distribution scaled to the mean of the powers in each bin. Here, we plot the standard deviation in each of the 5 Hz bins divided by the mean of the distribution for each bin. This quantity should be close to $1$ for pure red noise. The assumption holds above about 30 Hz, but becomes invalid below. Thus, statements for QPO features below 30 Hz should be interpreted with caution.}
\label{fig:michiel_dist}
\end{center}
\end{figure}

To make a first rough estimate of the effect of the relative strength of the burst envelope, we take all 27 bursts from the sample described in Section \ref{sec:datareduction} and stretch each light curve to have the same total length of $0.2$ seconds in order to make the time scales comparable. We then computed the Leahy-normalised periodogram for each light curve, gathering all powers in $5$ Hz-bins for all 27 bursts. This yields distributions of powers for each 5 Hz bin. For a noise process, the observed powers should follow a $\chi^2_2$ distribution scaled by the mean power in each bin. Thus, computing the standard deviation for each bin and dividing by the mean power should yield a value close to $1$, if the powers are truly $\chi^2_2$ distributed. This assumption may be broken in two possible ways. First, for steep power laws, the mean in a $5$ Hz bin may drop significantly, yielding powers that do not follow a $\chi^2_2$ distribution. Secondly, for bursts that vary significantly in brightness, the low-frequency red noise component may vary between bursts, and again, the distribution will be altered from our expectation. 
In Figure \ref{fig:michiel_dist}, we show exactly the dependence on frequency of this quantity. Above 30 Hz, the data seems to follow the noise distributions fairly well, while below 30 Hz, it deviates significantly upwards. There are multiple possible reasons for this. While we have corrected for the differences in burst durations, we have not normalized for the differences in fluence. Since burst fluences vary by over an order of magnitude within the sample, this may significantly increase the variation in burst periodograms. Alternatively, differences in burst envelope may account for some of this variation as well. It should be noted that one fundamental assumption underlying this test is the idea that all bursts are governed by the same kind of red noise spectrum. This may not necessarily be true, especially for bursts varying by over an order of magnitude in fluence, and a larger sample of bursts would be needed to draw any solid conclusions about the burst envelope from this kind of analysis. 
We conclude, for the purpose of our analysis, that the burst envelope seems to become largely unimportant above 30 Hz, and thus above this threshold our assumption of pure red noise is reasonable. Below, one should regard any conclusions drawn about QPOs with caution. However, this simple estimate is only provided to give an idea of where the burst envelope might be important. In the following, we perform detailed simulations of various kinds of light curves, both including and excluding a burst envelope, red noise and periodicities, in order to probe the effect the different components may have on the detectability of QPOs under the assumption of pure red noise. 
 
 \subsection{White Noise Simulations}
 
 \begin{figure}[]
\begin{center}
\includegraphics[width=9cm]{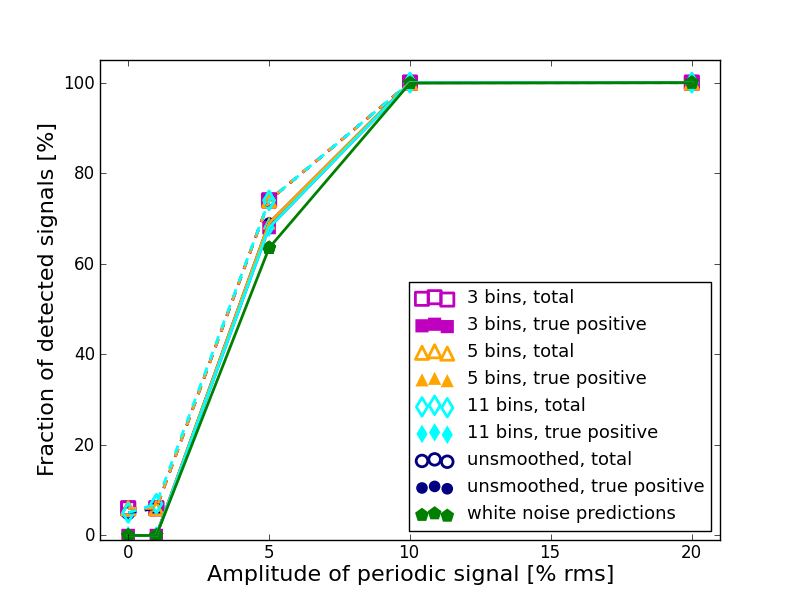}
\caption{Detection rates for simulated light curves of pure white noise (a light curve of constant count rate), a strictly periodic signal at 100 Hz and Poisson noise. We varied the fractional rms amplitude between 1\% and 20\% and compared with theoretical predictions calculated using the formalism in \citet{groth1975}. Different symbols and colours indicate (total) detection rates for the unsmoothed periodogram and three smoothing factors included in our analysis. Hollow markers and dashed lines correspond to total detection rates, filled markers and solid lines true positive indicate the number of true positive detections, defined here as the detections at the frequency where the signal was injected (as opposed to the total detection rate, which includes both true positive and false positive detections). Note that for white noise simulations, total and true positive detection rates practically lie on top of each other. 
True positive detection rates for our simulated light curves match white noise predictions (in green) within the uncertainties (not shown), indicating that our method performs equivalently well to standard Fourier techniques in the white noise regime. Since the rates of true positive detections trace the total detection rates fairly closely, we conclude that our method is not hampered by excessive numbers of false positive detections. For transient phenomena, this regime includes all frequencies above which slowly varying features in the light curve, e.g. red noise, do not dominate the power spectrum. }
\label{fig:whitenoise_detecrates}
\end{center}
\end{figure}

In order to test the detectability of (quasi-)periodic signals in complex burst light curves, we simulated a large number of fake observations of bursts and injected a periodic signal with varying frequency and fractional rms amplitude in order to cover a large range of possible signals. The phase of the injected periodic signal was randomized for all simulations to avoid correlations between simulations. For each combination of frequency and fractional rms amplitude, we simulated 100 light curves which we then ran through our analysis method as if they were real observations. While this is not enough to draw solid statistical conclusions about detectability rates, it gives a qualitative idea of what can be detected and what cannot. 

It is important to note that the amplitude of the signal we quote in all of this section is the fraction by which the underlying emission will be modulated. If the underlying signal is flat, then this amplitude corresponds to the fractional rms amplitude as measured from the periodogram. However, if the burst and the periodic signal vary together, such that the fractional amplitude at each point in the light curve is constant, this is not longer true. The reason for this discrepancy lies in the fact that a multiplicative process such as the one described here corresponds to a convolution in the Fourier domain, which will keep the product of the power in both processes - the burst and the periodic signal - constant, but redistributes power towards frequencies close to that of the periodic signal. The result will be a broadened peak in the power spectrum instead of a delta function. While the power in the two central bins will be the constant fractional rms amplitude corresponding to that we would have measured for al flat light curve with a periodicity, the side wings due to the convolution supply power that may be, in practice, indistinguishable from an intrinsically broad QPO. Hence, one would include these side bands into the calculation for the fractional rms amplitude, and in practice measure an amplitude that is larger than that we put in. A characterization of this effect is beyond the scope of this paper; we merely wish to remind the reader that they must take these effects into account when considering the fractional rms amplitudes quoted in this Section. 

In a first step, we tested the simplest case: (flat) white noise. We created flat light curves with a constant count rate, a periodic signal at 100 Hz of varying fractional rms amplitude, randomized phase and Poisson noise. In this limit, our method should match standard Fourier analysis techniques and follow the predictions of \citet{groth1975}. In Figure \ref{fig:whitenoise_detecrates}, we show the theoretical predictions for white noise together with the results of our simulations. The observed detection rates match the white noise predictions fairly well, and our method deviates from expectations only for a fractional rms amplitude of $5\%$, but remains within the uncertainty (a 5\% error on a detection rate of $\sim 0.6$ is expected based on 100 simulations). Thus, in the limit of white noise, our method is equivalent to standard Fourier techniques. For any signal at higher frequencies, where slowly varying processes do not distort the power spectrum, we will be as sensitive to a QPO as standard techniques. It should be stressed that the probabilities we quote include a {\it Bonferroni} correction for the number of frequencies, and are thus not directly comparable to single-trial detection probabilities. \\

\begin{figure}[htbp]
\begin{center}
\includegraphics[width=9cm]{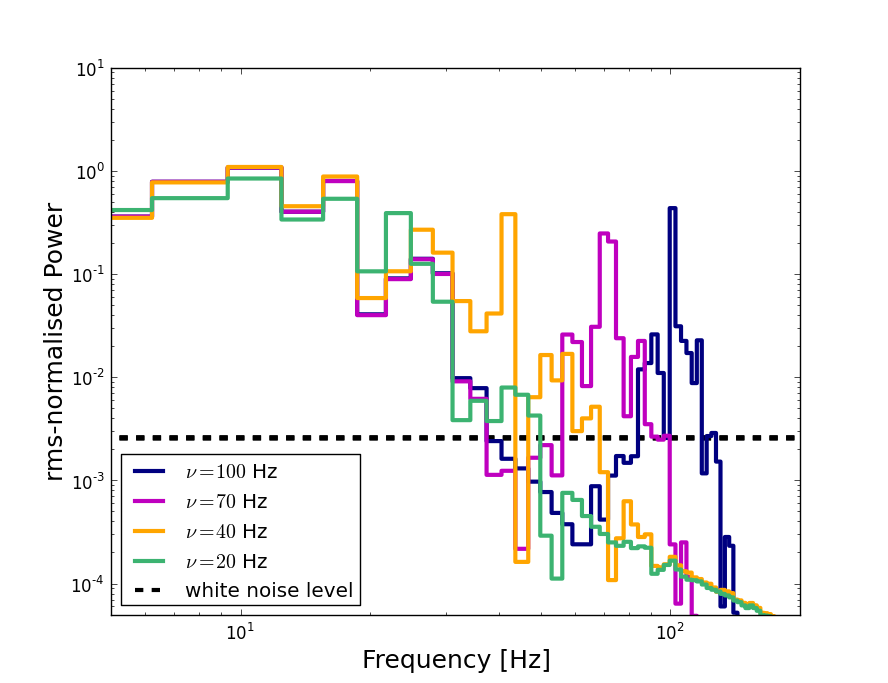}
\caption{Periodograms of simulated light curves including a complex burst envelope shape and a periodic signal, without Poisson noise. We varied the frequency from 20 Hz to 100 Hz, and kept a constant fractional rms amplitude of 20\%. The signal at 20 Hz (green) is almost invisible, and likely impossible to distinguish from the underlying burst envelope, hence we do not consider signals this low in the following analysis. For a signal at 40 Hz (orange), our method is unlikely to be able to distinguish between the broadened periodic signal and the burst envelope, causing low detection rates for even high rms amplitudes. As the signal moves to higher frequencies (magenta: 70 Hz, dark blue: 100 Hz), detection rates converge towards the detection rates predicted for white noise (black dashed line).}
\label{fig:oldmc_signals}
\end{center}
\end{figure}

\subsection{Pure Burst Envelope Simulations}
\label{sec:pureburst_sims}

In order to test the effect of a burst envelope on detectability, we started with the extreme assumption: the burst is dominated by a complex burst envelope and Poisson statistics, lacking any red noise. In order to generate the complex burst envelope, we smoothed the light curve of {\it 080823478} (see Figure \ref{fig:bmexmp_comb}) to an arbitrary cut-off frequency, in our case roughly $35 \, \mathrm{Hz}$, creating a smooth light curve with several broad peaks. We included Poisson noise in each simulation, and a periodic signal at $40, 70$ and $100 \, \mathrm{Hz}$ with an absolute amplitude that varied with the flux of the burst envelope such that the fractional root-mean-square amplitude remained constant. We varied the fractional rms amplitude between $1\%$ and $20\%$, and ran simulations without a periodic signal in order to quantify false positive detection rates.

In Figure \ref{fig:oldmc_signals} we present a selection of periodograms of the resulting combined light curves without Poisson noise. Most notably, the multiplication of a complex burst envelope with a periodic signal in the light curve leads to a significant broadening of the periodic signal in the Fourier domain, including wings and side-lobes. 
In this scenario, signals below $40 \, \mathrm{Hz}$ should be undetectable, whereas at higher frequencies detection rates should approach what we would predict for pure white noise. The combination of the envelope from the fit solution and a periodic feature additionally changes the slope of what our method will interpret as broadband noise in the case where a periodic signal is located just at the break where the noise powers drop towards the white noise level. We predict that this will lead to decreased detection rates as well.

\begin{figure}[htbp]
\begin{center}
\includegraphics[width=9cm]{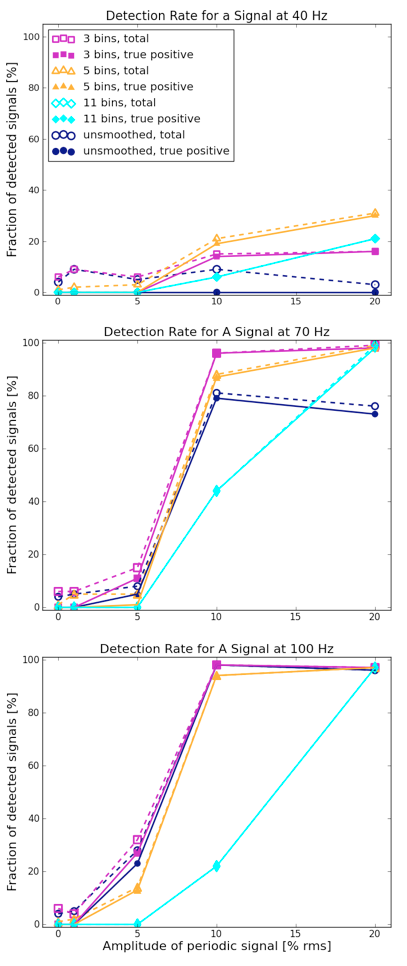}
\caption{Detection rates for periodic signals at various frequencies on top of a smoothed burst envelope. The envelope is generated by smoothing the light curve of burst {\it 080823478}. This introduces a sharp cut-off at around 35 Hz (corresponding to the timescale of smoothing) that is unlikely to be seen in real light curves. Detection rates at 40 Hz are strongly suppressed, indicating that at low frequencies, chances of detecting even a very strong signal are small. Note that while there are no true positive detections (as defined in the caption of Figure \ref{fig:whitenoise_detecrates}) at 40 Hz for the unsmoothed periodograms, the number of true positives for the binned periodograms closely traces the total detection rates. Detection rates increase with frequency and fractional rms amplitude as expected.}
\label{fig:oldmc_detecrates}
\end{center}
\end{figure}

\begin{figure*}
\begin{center}
\includegraphics[width=18cm]{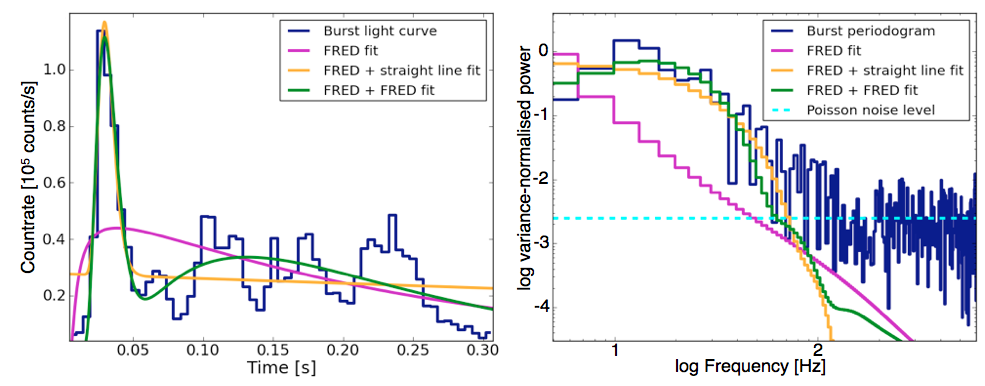}
\caption{This figure shows the effect that different choices of burst envelope have on conclusions for the relative strengths of the red noise component and envelope component in the low-frequency part of the periodogram. Left: light curve of burst 080823478 with single-component fast-rise exponential decays (FRED) fit (magenta), a FRED profile with a straight line to account for the long tail (orange) and a combination of two FRED profiles (green); right: periodogram of the same burst and all three fits from $4$ to $600 \, \mathrm{Hz}$. There are clear differences in how strong the envelope is at low frequencies. The Poisson noise level is shown in a light blue dashed line for comparison.}
\label{fig:benv_1}
\end{center}
\end{figure*}

Figure \ref{fig:oldmc_detecrates} presents total detection rates as well as true positive detection rates (both out of 100 simulated bursts) at $40, 70$ and $100$ Hz for five different fractional rms amplitudes and both unsmoothed and smoothed periodograms. True positive detection rates are measured as detections at the frequency of the injected periodic signal. 
As predicted, detection rates increase towards higher frequencies, where the envelope becomes unimportant. 
While there are no true positive detections for the unsmoothed periodograms at low frequencies, the number of true positives for the smoothed periodograms closely traces the total detection rates. This points towards a significant broadening of the periodic signal as a result of its convolution with the burst envelope. In general, at frequencies below 100 Hz, detection rates for periodograms smoothed to three or five bins are higher than for the unsmoothed periodograms or those smoothed over 11 bins. At higher frequencies, the detection rates for unsmoothed periodograms approach those of the smoothed periodograms. 
At 5\% fractional rms amplitude or below, there may be a significant contribution from false positive detections, which vanishes for higher fractional rms amplitudes and higher frequencies.

\subsection{Envelope Plus Red Noise Simulations}
\label{sec:envrednoisesims}

\begin{figure}
\begin{center}
\includegraphics[width=7cm]{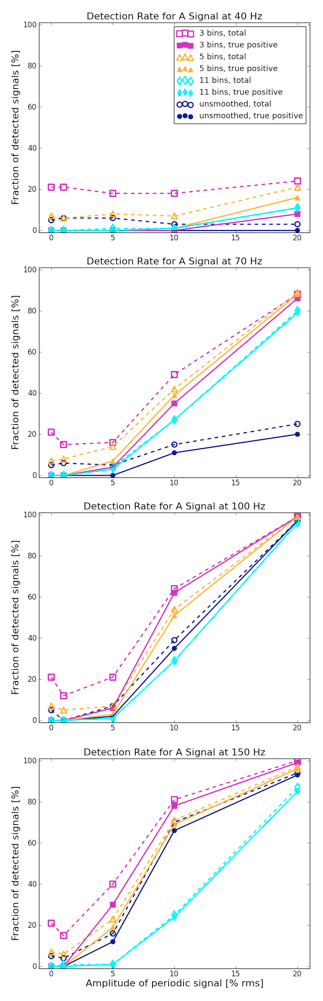}
\caption{Total (hollow markers and dashed lines) as well as true positive (filled markers and solid lines; for a definition see caption of Figure \ref{fig:whitenoise_detecrates}) detection rates for a periodic signal on top of a combined burst envelope and red noise light curve. The different panels correspond to different frequencies of the periodic signals, from $40$ to $150$ Hz. As before, we include detection rates of both the unsmoothed and smoothed periodograms. Note that for the case where a periodic signal is combined with a burst envelope and red noise, the periodic signal will be significantly broadened, and thus one can more successfully detect these signals in binned or smoothed periodograms. This is especially true at 70 Hz. For higher frequencies, the difference becomes smaller, but is still appreciable.}
\label{fig:envrednoise_detecrates}
\end{center}
\end{figure}

We used light curves composed of both a burst envelope with a simple functional form as well as a power law red noise component, combined with Poisson statistics and a periodic signal of varying frequency, amplitude and randomized phase. Although both the burst envelope and the red noise shape are guesses and to some degree degenerate - a different choice of burst envelope shape may lead to a different estimate of the red noise power spectrum - we believe that this type of light curve is likely to be more realistic than the pure burst envelope model, based on the observation that bursts consistently rise faster than they decay (indicating the presence of a deterministic component) and the large variety in burst shapes otherwise (indicating the presence of some form of variability on many time scales that is typical for red noise).

For the largely qualitative conclusions we wish to draw here, neither the exact shape of the burst envelope, nor the exact parameters of the red noise are important, although both are interesting questions in their own right and beyond the scope of this paper. Instead, we wish to give a representative example of the general behaviour one may expect when applying the method presented here to light curves of transient events with complex light curves.

As before, we use burst {\it 080823478} as a template burst on which to base our simulations.
This burst presents an interesting profile, with a main spike and several features that are reminiscent of red noise (see Figure \ref{fig:benv_1}). We cannot exclude the possibility that the latter is actually due to a more complicated emission mechanism, and can only state that its timing properties are consistent with red noise. 
We fit the entire light curve with several models, all based on a fast-rise, exponential decay (FRED) profile of the type

\begin{equation}
\label{eqn:fred}
f(x) = A \lambda \exp\left(\frac{-\tau_1}{(t-t_s)} - \frac{(t-t_s)}{\tau_2}\right) \; .
\end{equation}

Here, $\tau_1$ and $\tau_2$ are the rise and decay timescales, respectively, $t_s$ is the burst start time, $A$  is a normalization constant (or burst amplitude) and $\lambda = \exp(2(\tau_1/\tau_2))^{1/2}$ \citep{2010ApJ...718..894P}. This model has been successfully applied to gamma-ray bursts (GRBs) in the past and appears to be a reasonable first assumption for magnetar bursts with their exponential-like tails and shorter burst rise times compared to the decay timescale. Because the burst has a sharp initial spike and then a long, relatively flat, but very variable tail, a single FRED profile has trouble fitting the entire light curve well: it can either fit the main spike with its sharp decay, or the long tail, but not both together. Hence, we implement two more complex hypotheses: a FRED profile to account for the initial spike, on top of a linear function modeling the slow decay, as well as a model with two FRED components. The former does not fit the beginning and end of the burst well, as it does not drop to background noise level as it should at the start and end of the burst. The latter provides the best fit of the three, but is the model with the largest number of parameters and requires an explanation for the origin of the additional FRED component. 
The periodogram presented in the right panel of Figure \ref{fig:benv_1} shows how important the choice of burst envelope is for disentangling red noise and deterministic envelope at low frequencies: if the burst envelope could be modeled with a single FRED profile, the low-frequency part of the power spectrum would be entirely dominated by red noise, and the assumptions we make in Section \ref{sec:bayes} hold to a fairly high degree. If the model has additional components, however, either in form of a straight line, another FRED profile or another type, this component will dominate the periodogram up to about $60$ or $70 \, \mathrm{Hz}$. As a consequence, assuming pure red noise in this part, if the more complex hypothesis were true, our assumption of pure red noise might be a poor one in this frequency range. 
For what follows, we choose the combination of FRED and linear model, to keep our model as simple and the number of free parameters as low as possible.

Having chosen a model for the overall burst morphology, we make an estimate of the red noise part of the power spectrum: we de-trend the light curve by dividing the light curve by the lightcurve model fit and compute the periodogram of the residuals. De-trending in this way will give us a light curve fluctuating around a mean of $1$, and in line with our assumption, we consider the variance around that mean to be red and white noise only. We fit a power law (for the red noise) plus a constant (accounting for Poisson noise) to the periodogram of the residuals, and take the resulting power law fit as a template power spectrum to simulate red noise light curves from.
Using the method from \citet{1995A&A...300..707T}, we simulate $100$ light curve realizations from red noise power spectra only. Note that light curves simulated according to \citet{1995A&A...300..707T} will have entirely uncorrelated phases, which may introduce a bias into the light curves if this does not accurately reflect reality. More importantly, light curves generated this way are distributed around a mean of zero. In reality, light curves with negative count rates are unphysical, however, any transformations applied to the simulated light curve will result in correlations between phases in the periodogram. 
We choose a method following \citet{2005MNRAS.359..345U} to generate log-normally distributed light curves that have no negative data points, accepting that the assumption of log-normally distributed light curves introduces a potential bias into our simulations via the correlations it introduces between the phases in the periodogram. 
Each red noise light curve is combined with the template assumed for the burst envelope. This will provide us with $100$ light curves including both an envelope and a red noise component which we can use as fake observations to be analysed through our method. 
We add periodic signals in the same way as in Section \ref{sec:pureburst_sims}, however, since the red noise we included in the simulations does not drop off as sharply as the burst envelope in the previous section, the periodogram at $100$ Hz is still contaminated by red noise. Thus, here and in the following section, we also include simulations with a periodic signal at $150$ Hz to probe the white-noise dominated region, and run each light curve through our Bayesian detection method.

The detection rates for various frequencies are shown in Figure \ref{fig:envrednoise_detecrates}. Signals at 40 Hz are not detectable, with either no or very few true positive detections. Detection rates rise for higher frequencies towards the white noise limit, although even for $150$ Hz, a signal at a fractional rms amplitude of 5 \% or 10 \% is still somewhat suppressed. The effect of red noise on detectability is more wide-spread in frequency compared to the case of a pure burst envelope, where the power due to broadband variability drops sharply around 40 Hz. Binned or smoothed periodograms are generally better at detecting periodic signals combined with a burst envelope and red noise, with detection rates in the unsmoothed periodogram only approaching the performance of the smoothed spectra (and at the same time the white noise limit) for high frequencies. This again is due to the broadening of the periodic signal in the convolution with the burst envelope and red noise. As for the pure burst envelope simulations, total detection rates contain a significant contribution from false positive detections, and true positive detection rates approach total detection rates for high frequencies and large fractional rms amplitudes. We note that the false positive detection rate at low fractional rms amplitude seems uncharacteristically high for the 3-bin periodogram at all frequencies in these simulations. At present, we do not understand the reason for this. It is possible that the broadband fitting is not entirely reliable in some of the simulations. In practice, the results from the broadband fitting of the periodograms of real bursts are checked to ensure minimization to a global minimum. Additionally, we believe the number of false positive detections is easily corrected for by requiring that, in practice, signals at low fractional rms amplitude need to be significant in at least two different smoothed or binned periodograms.

\subsection{Pure Red Noise}
\begin{figure}
\begin{center}
\includegraphics[width=6.5cm]{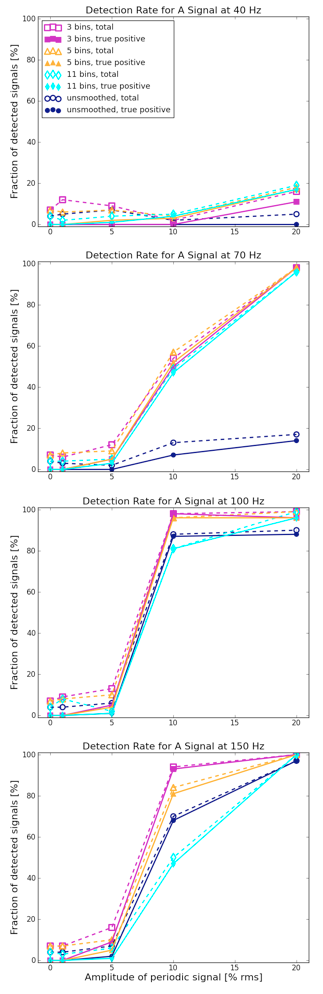}
\caption{Detection rates for a periodic signal on top of a red noise light curve. The different panels correspond to different frequencies of the periodic signals, from $40$ to $150$ Hz. As before, we include total (hollow markers and dashed lines) and true positive (filled markers and solid lines, definition in caption of Figure \ref{fig:whitenoise_detecrates}) detection rates of both the unsmoothed and smoothed periodograms. Note that signals at 5\% rms amplitude or below are generally suppressed, as compared to the white noise detection limit accessible with standard Fourier techniques.}
\label{fig:purerednoise_detecrates}
\end{center}
\end{figure}

As a last example, we test detectability under the hypothesis that our light curve has no deterministic element at all, and consists purely of red noise. We generate light curves using the method from \citet{1995A&A...300..707T}, using the fit to the periodogram of burst {\it 080823478} as a template to achieve comparable burst length, fluence and rms variability. Again, we introduced a periodic signal of constant rms amplitude and randomized phase into each light curve, changing the fractional rms amplitude of the signal for different simulations. Figure \ref{fig:purerednoise_detecrates} presents the detection rates for the simulations of pure red noise.
Detection rates for the pure red noise case are comparable to the case where there is an envelope component, indicating that the presence of a burst envelope does not significantly alter detectability of a periodic signal if there is a red noise component present. Low-amplitude signals are suppressed in the case where the light curve consists only of red noise compared to simulations including a burst envelope. As for the combined light curves, detection rates for one or more of the binned periodograms are always equal or higher than for the unbinned periodograms, and below $70$ Hz, even strong signals become nearly undetectable.  
For high frequencies, the detection rates approach the white noise predictions and are slightly higher than predictions for simulations including both a burst envelope and a red noise component, although detection rates for a fractional rms amplitude of $5 \%$ are suppressed even for a $150$ Hz periodic signal. This indicates that the red noise component is still significant at these frequencies, for the given input spectrum. In reality, the frequency at which the white noise detection limit holds will depend on the specific burst variability for each observation.

\subsection{Conclusions from the Simulations}

The different types of simulations allow us to draw two important conclusions for QPO detection: (i) in the limit of a flat light curve, translating to a pure white noise power spectrum, our method does equally well compared to standard Fourier techniques, and (ii) detection rates for more complex light curves depend on the underlying emission mechanism. Even a periodic signal may be significantly altered (i.e. broadened) by the presence of a burst envelope and/or red noise, if the periodic signal is modulated by these aperiodic processes, and this broadening will depend specifically on the shape of burst envelope and the red noise parameters, as well as the relative strengths between the two. A significant broadening may in turn affect detectability when it alters what our method interprets as broadband noise, decreasing detection rates even for high fractional rms amplitudes.
For the small sample of bursts from SGR J0501+4516, a simple, crude estimate comparing the standard deviation to the mean in small frequency bins across all bursts in the sample reveals that the assumption of red noise holds reasonably well for frequencies above 30 Hz. Below, the assumption may either be broken by the presence of a burst envelope, or alternatively the bursts may be sufficiently varied in red noise properties to produce the observed increase in standard deviation about the mean. This need not mean that our assumption of red noise is invalid in this regime, simply that we do not know this to be true or false. Hence, we caution the reader to interpret results at frequencies this low with care and with the conclusions of the burst simulations in mind.

In general, signals below 70 Hz or so will be very difficult to detect, unless they have fractional rms amplitudes of above $10 \%$. This is not impossible, given the high fractional rms amplitudes observed from the 2004 giant flare (see \citealt{Watts11}, Table 1 for an overview), however, even for high amplitudes false non-detections may still occur. We recognize these issues as a shortcoming of the presented method, however, in the absence of any physical model or empirical evidence for a consistent burst envelope structure, we opt for the conservative approach presented here. Thus, we caution the reader to keep the effects described above in mind when interpreting the posterior p-values and sensitivities quoted in Section \ref{sec:results} below. 

On the other hand, we have also shown that while the sensitivity of our method depends on the type of light curve analysed, detection rates for both light curves combining an envelope and red noise - the case we consider most likely for SGR bursts - and for pure red noise light curves, detection rates are quite similar, within the uncertainties, indicating that the additional envelope component does not significantly alter our chances of detecting a signal. Hence, unless light curves are purely deterministic, our method will yield fairly reliable results. At the same time, the false positive detection rate is generally low in most simulations, which is one of the key goals of developing this technique for transients. False positive detections can be dealt with by requiring detection in more than one smoothed or binned periodogram.

Finally, we would like to make two notes: First, the findings above are based on the assumption that a periodic signal will vary with the underlying light curve, that is, that the fractional rms amplitude, rather than the absolute amplitude, remains constant. This assumption, of course, need not be true. Instead, the absolute amplitude may be constant, in which case a periodic signal would truly remain confined to two frequency bins, and none of the broadening would occur. 
Secondly, we also note that the false positive detection rate is low, as expected for a conservative approach. For 100 fake observations, and a detection threshold of $p<0.05$ for each observation, we find roughly 5 false positive detections in most runs, exactly as expected. The sole exception is the run that combines a burst envelope and red noise. At present, it is not clear what causes this increase in false positive detection rate. The false positive detection rate can be lowered by tightening the detection threshold to a smaller probability, at the cost of increasing the upper limit to the amplitude of a signal we might have missed, or, in other words, increasing the risk of false non-detections.

\section{Results}
\label{sec:results}

We computed light curves and periodograms for all 27 bursts (Sections \ref{sec:exampleburst} and \ref{sec:allresults}) as well as time segments before and after each burst (Section \ref{sec:beforeafter}). In each case, we produced a light curve by binning the TTE data to a time resolution of $1/2\nu_{\mathrm{Nyquist}} = 1.22\times 10^{-4} \, \mathrm{s}$, corresponding to a Nyquist frequency of $\nu_{\mathrm{Nyquist}} = 4096 \, \mathrm{Hz}$. The time resolution may be arbitrarily chosen, as long as it remains poorer than the time resolution of the detector itself, i.e. $2\, \mu \mathrm{s}$ for Fermi GBM, although searches with high frequency resolution up to large Nyquist frequencies quickly become computationally expensive. We chose the time resolution based on the Nyquist frequency of interest: we do not expect any signals above $4000 \, \mathrm{Hz}$ from neutron star seismic oscillations \citep{1988ApJ...325..725M}. 
We search both the unbinned periodogram as well as the same periodogram binned to integer multiples (3, 5, 7, 10, 15, 20, 30, 70, 100, 200, 300, 500 and 700) of the frequency resolution of that burst, i.e the actual frequency resolution of the binned periodograms depends on the frequency resolution of the unbinned periodogram. Additionally, we smooth the spectra with a Wiener filter with different smoothing factors (3,5, and 11) and compare results of the search of binned periodograms with searches across the smoothed periodograms. Note that while computing sensitivities for binned periodograms is statistically straightforward, doing so for a convolution of the periodogram and a smoothing function is not, hence all sensitivities quoted refer to either the unbinned or binned periodogram, but never the smoothed one.  

\subsection{Checking for spurious timing signals}
\label{sec:beforeafter}
Fermi GBM sees the entire unocculted sky at any given point in time. Therefore, the $\gamma$-ray background can be rather complex, and one must exclude that a background source supplies significant variability to the burst periodogram. To this end, we performed timing analysis on $1 \, \mathrm{s}$ and $10 \, \mathrm{s}$ long segments before and after each burst as well as on the bursts themselves. The light curves constructed out of these segments were  Fourier transformed and normalized in order to produce periodograms with a Leahy normalization \citep[noise powers averaging to $2$, ][]{1983ApJ...266..160L}. 

For none of the segments before and after each of the $27$ bursts in our sample did we find significant detections of periodicities or QPOs. All segments present white-noise dominated periodograms consistent with a Poisson noise $\chi^2$ distribution, indicating that the burst emission is not contaminated by a background source with significant timing behaviour or instrumental effects on the relevant time scales. This includes any potential signal from the source itself. Any additional background source contributing emission would have to have switched on at the same time as the burst occurred, and switched off equally fast. This is highly unlikely.
Note, however, that some instrumental effects, particularly dead time, scale with the source flux, and will not be recognizable in the background periodograms. Dead time in particular has an intricate effect on the burst periodogram, and led us to exclude the brightest bursts, which were also saturated.

\subsection{An Example:  Timing Analysis of Burst 080823478}
\label{sec:exampleburst}
\begin{figure}[hbtp]
\begin{center}
\includegraphics[width=9cm]{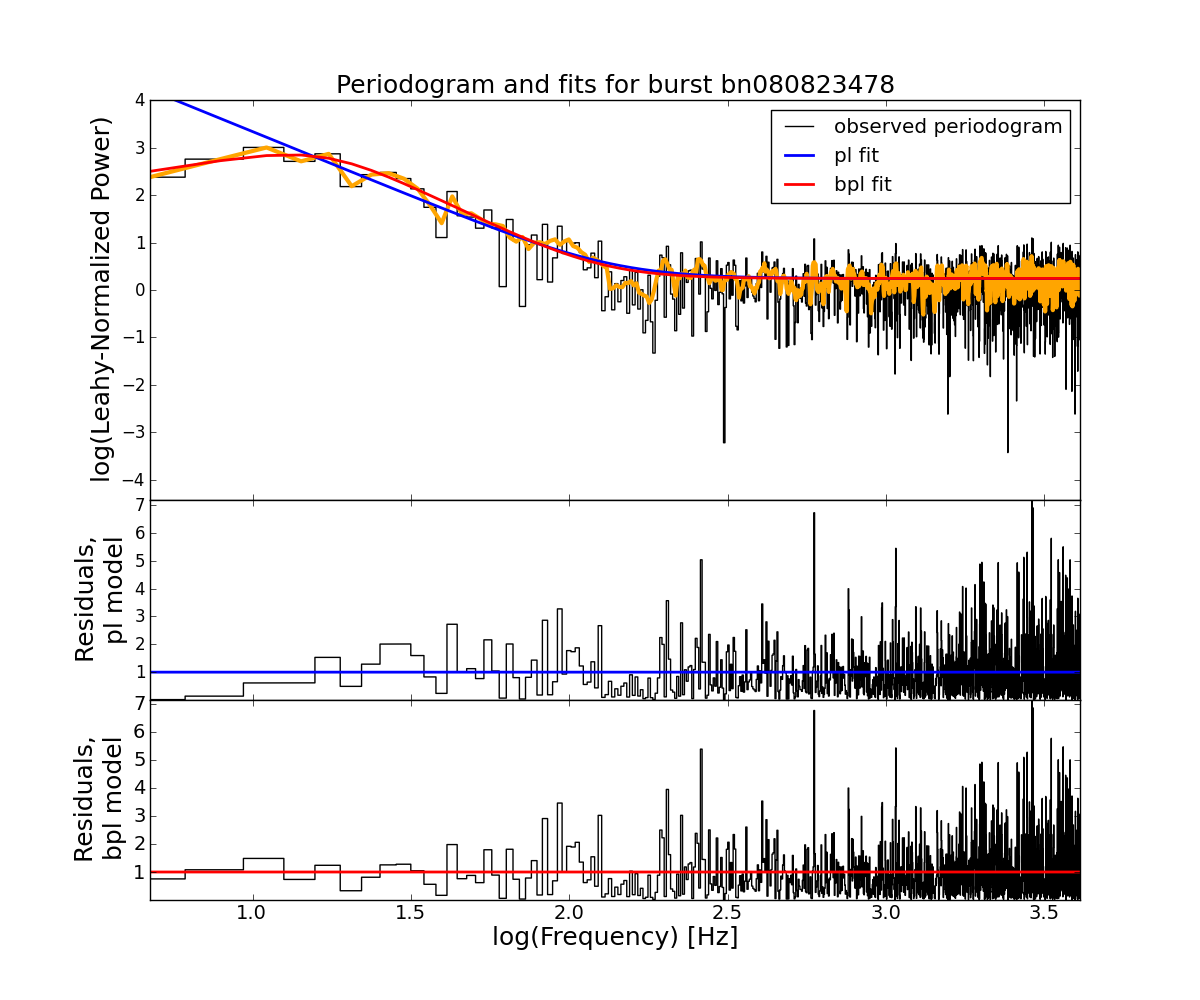}
\caption{Fermi GBM observation of burst {\it bn080823478} from SGR J$0501+4516$: periodogram and residuals $I_j/S_j$ (for the light curve, see Figures \ref{fig:bmexmp_comb} and \ref{fig:benv_1}). Upper panel: unsmoothed (black) and smoothed (orange; Wiener filter, $5\Delta\nu$) periodogram, power law fit (blue) and broken power law fit (red). Middle panel and lower panel show the residuals of the power law fit and broken power law fit, respectively. The broken power law presents a significantly better fit to the data.}
\label{fig:bexmp_lc}
\end{center}
\end{figure}

In the following, we illustrate the analysis procedure with one specific burst, {\it  bn080823478}, before giving results for the whole sample.  This burst had a duration of $T90 = 264 \, \mathrm{ms}$ and the highest fluence of the sample (see Table \ref{tab:bayes_results}). The periodogram for this burst is presented in Figure \ref{fig:bexmp_lc}.

\subsubsection{Choosing a Noise Model}
\label{sec:bn080823478}
\begin{figure*}
\begin{center}
\includegraphics[width=18cm]{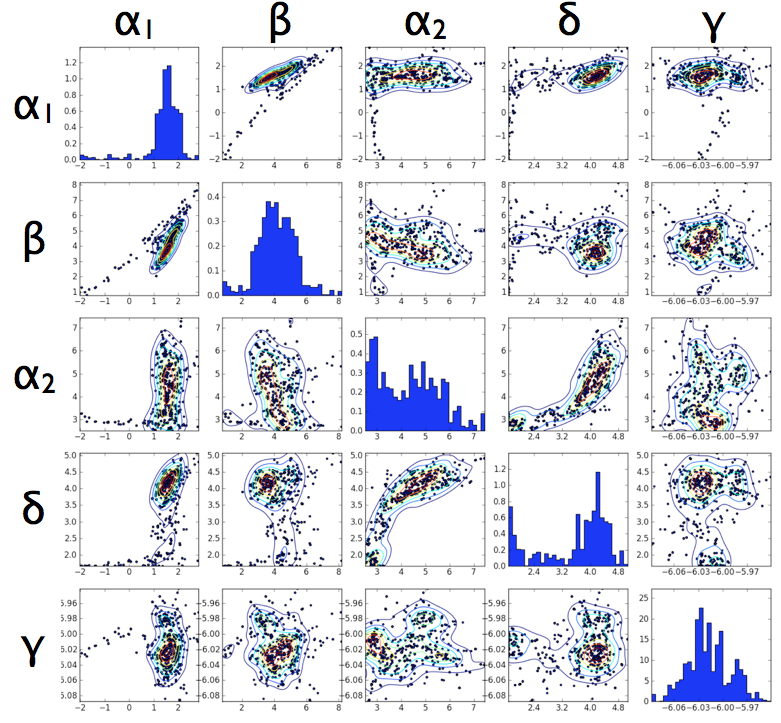}
\caption{The MCMC sample of the parameters for the preferred broadband model for burst {\it 080823478} (here: a broken power law). The posterior distributions for individual parameters are presented on the diagonal. If a posterior distribution is very broad, then the parameter is not very well constrained (indicating a high standard deviation on that parameter), and a simpler model might be adequate. The off-diagonal panels show correlations between parameters (panels opposite of each other, mirrored on the diagonal, are equivalent): scatter plots for 1000 randomly picked parameter pairs from the entire sample of 250 000 parameter sets, and contours of number density. One can observe for example a very tight correlation between low-frequency power law index and normalization, and very little correlation between the normalization and the noise. Other parameters may correlate in more complex ways with each other. The trails and ``clumpiness'' in some of the scatter plots indicate that the distributions are not perfectly unimodal, and that even for highly peaked distributions, there are parameters far off the mean that are nevertheless not entirely unlikely.}
\label{fig:bexmp_scatter}
\end{center}
\end{figure*}

\begin{figure}[h]
\begin{center}
\includegraphics[width=8cm]{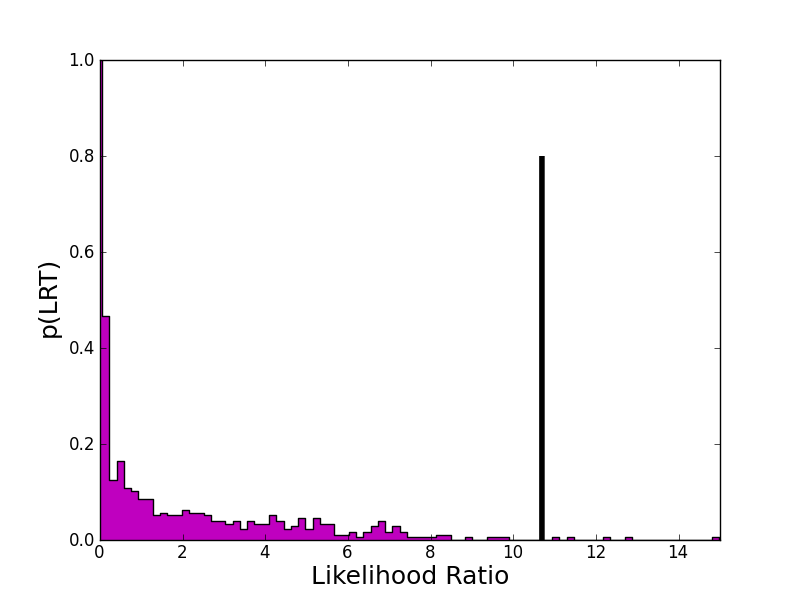}
\caption{Distribution of likelihood ratios for 1000 simulations of the null hypothesis (power law model). The observed value of $T_{\mathrm{LRT}}$ for burst {\it 080823478} is indicated as a black vertical line. The further to the right (i.e. the further in the tail of the distribution) this observed value is located, the more unlikely the null hypothesis becomes, indicating that a more complex model (in this case the broken power law) may be more appropriate in modeling the broadband variability. }
\label{fig:bexmp_lrt}
\end{center}
\end{figure}

After fitting both a simple power law and the more complex broken power law, we computed the likelihood ratio between the two models, $\mathrm{LRT} = 10.69$. Note that here, as well as in the analysis of the remaining sample, we set the smoothness parameter of the broken power law to $-1$ as in \citet{vaughan2010}. The resulting function should more correctly be called a bending power law in this case, since it turns over in a smooth bend rather than a sharp break. Setting the smoothness parameter to $-1$ introduces a potential bias into the determination of the low-frequency power law index for this model, however, including the smoothness parameter in the MCMCs, we found that the posterior distributions of this parameter for the bursts in our sample are very broad, indicating that the parameter is unconstrained. At the same time, it is correlated with the low-frequency power-law index, and thus the quoted values for this parameter should be read with caution. Additionally, we show below that the overall goodness of fit of the model to the data is good, indicating that another component is not needed.
The fits to the periodogram and the residuals (data/fit) are presented in Figure \ref{fig:bexmp_lc}. 
We use the Gaussian approximation to the covariance and the best-fit model parameters for the power law model ($H_0$) as input to 500 MCMC ensemble walkers (see Section \ref{sec:bayes} or \citealt{2012arXiv1202.3665F} for details) with 100 samples each, after a burn-in phase with $100$ samples for each walker. Figure \ref{fig:bexmp_scatter} presents the posterior distributions of all five parameters and their correlations with each other.
With 1000 randomly picked parameter sets from this sample of 50000 parameter sets, we create 1000 fake periodograms, and compute the posterior predictive p-value for the LRT of the observed data using the formalism outlined in Section \ref{sec:bayes}. In Figure \ref{fig:bexmp_lrt} we plot a histogram of the posterior distribution for the likelihood ratio from the simulated periodograms. The black vertical line indicates the value of the likelihood ratio of the observed data.
For bn080823478, $p(LRT) = 0.003 \pm 0.002$, hence we consider observing these data unlikely under the null hypothesis (a simple power law), and we adopt model $H_1$ for the rest of our analysis of this burst.

\subsubsection{Searching for Periodicities}

\begin{figure}[h!]
\includegraphics[width=9cm]{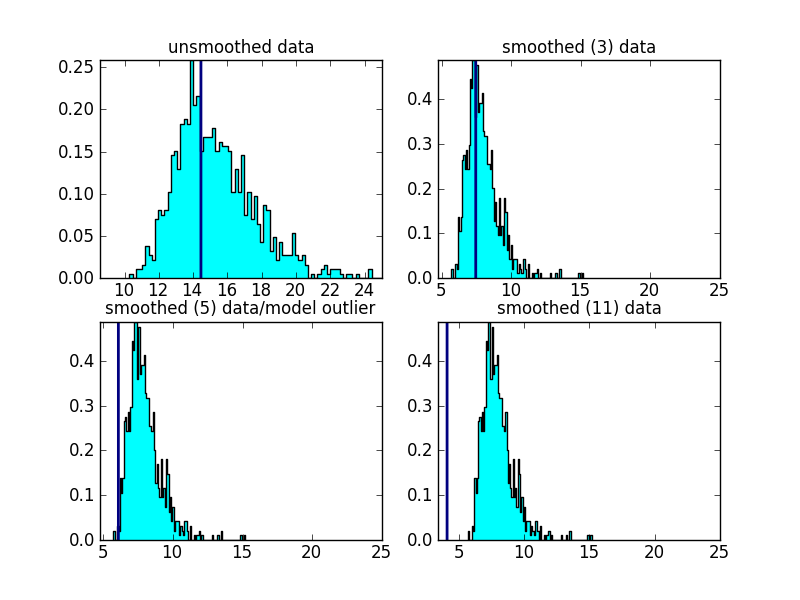}
\caption{Histograms of posterior distributions and observed values (black vertical lines, burst {\it 080823478}) of the $T_R$ statistic defined in Equation \ref{eqn:trstat}.}
\label{fig:bexmp_maxpow}
\end{figure}

We  use the broken power law fit to the periodogram to draw another sample of MCMC parameter sets, in the same fashion as outlined above, and simulate 1000 fake periodograms all following a broken power law. From these, we computed the summed-square residuals $T_{\mathrm{SSE}}$ and search for the highest data/model outlier, $T_R$ in the unbinned and binned periodogram. 
The latter should tell us about any features narrower than the frequency resolution $n\Delta\nu$ (where $n=1$ for the unbinned periodogram and $n>1$ for binned periodograms), while the former will give information about the overall fit of the model to the data. 
For this burst, we computed the posterior predictive distribution for the square-summed residuals and compared this distribution to the observed value, finding $p_{\mathrm{SSE}} = 0.49 \pm 0.01$. As this statistic is an indicator for how well the model fits the data, we expect a low $p_{SSE}$ to indicate that the model fit could be improved, either by implementation of a different model or addition of model components. For this burst, we conclude the model fits the data rather well.
The highest outlier in the residuals is at $ \nu_{\mathrm{max}} = 2317 \, \mathrm{Hz}$ with a power $2I/S = 15.71$ and a posterior predictive p-value $p(T_R) = 0.42 \pm 0.02$. The observed maximum power seen in the residuals is well within the distribution of outliers produced by the Monte Carlo simulations of the broadband model without any periodicity (see Figure \ref{fig:bexmp_maxpow}), and is hence unlikely to represent a real periodic process. Similarly, we find maximum outliers as well as posterior probabilities of these outliers for the smoothed and binned periodograms, which show no significant features, either. The results are summarized with the remaining model parameters as well as the other bursts in Tables \ref{tab:bayes_results} and \ref{tab:findper_results}. None of the outliers were significant, thus we conclude that under our assumption of red noise, there are no narrow (quasi-)periodic signals in this data set. The posterior distributions and the observed values for the unsmoothed and smoothed periodograms are presented in Figure \ref{fig:bexmp_maxpow}. 

We searched the burst for broader quasi-periodic signals using an additional Lorentzian component and comparing the mixture model of broadband noise process and Lorentzian to the pure broadband model. With a posterior probability of the pure broadband model of $p(LRT) =  0.51 \pm 0.02$ (i.e. the probability that this model is sufficient in explaining the observed data), we conclude that there is no QPO in the burst.

\subsection{Whole Sample}
\label{sec:allresults}

\begin{deluxetable*}{|l|r|r|r|l|l|l|l|}
\tablewidth{500pt}
\tablecolumns{8}
\tablecaption{Posterior summary of results for the broadband modeling for all bursts in the sample. }
\tablehead{\colhead{Burst ID} &
\colhead{Length [ms]} &
\colhead{Fluence [$\mathrm{erg} \, \mathrm{cm}^{-2}$]} &
\colhead{Model} &
\colhead{$p(LRT)$} &
\colhead{$p(SSE)$} &
\colhead{$\alpha_1$} &
 \colhead{$\alpha_2$}}
 \startdata
080822529  & 86   &   7.05  &	flat & 0.31 $\pm$ 0.01 & 0.82 $\pm$ 0.01 & \nodata &\nodata  \\
080822647  & 216   &  19.3 &   	PL &  0.10 $\pm$ 0.01& 0.70 $\pm$ 0.01 & $2.41^{+0.39}_{-0.35}$ & \nodata   \\
080822981  & 30    &  4.41    &	PL  & 0.16 $\pm$ 0.01 & 0.84 $\pm$ 0.01& $2.42^{+1.44}_{-1.23}$  & \nodata  \\
080823020  &  66   &   25.02   & PL & 0.99 $\pm$ 0.002 & 0.55 $\pm$ 0.02 & $2.72^{}_{-0.65}$  & \nodata \\
080823091 & 676   &  82.84 & flat & 0.59 $\pm$ 0.01 & 0.02 $\pm$ 0.005 &\nodata & \nodata \\
080823174 & 447   &  14.3    & PL & 0.09 $\pm$  0.008 & 0.82 $\pm$  0.01 & $1.93^{+0.91}_{-0.71}$  & \nodata   \\
080823248   & 272 &    22.18  & PL & 0.29 $\pm$ 0.01 & 0.85 $\pm$ 0.01 & $4.19^{+1.95}_{-1.50}$ & \nodata    \\
080823293a  & 189  &   20.10 & PL & 0.11 $\pm$ 0.01 & 0.75 $\pm$ 0.01 & $2.65^{+0.61}_{-0.60}$ & \nodata   \\
080823293b  & 38    &  9.54  & flat &  0.09 $\pm$ 0.009 & 0.95 $\pm$ 0.006 & \nodata & \nodata  \\
080823319 & 142   &  19.42  & PL & 0.16 $\pm$  0.01 & 0.78 $\pm$  0.01 & $2.79^{+1.04}_{-0.70}$  & \nodata  \\
080823330   & 192 &    67.05 & PL & 0.47 $\pm$  0.02 & 0.18 $\pm$  0.01 & $2.71^{+0.36}_{-0.34}$ &  \nodata \\ 
080823354   & 96   &   8.62    &	 PL  & 0.51  $\pm$ 0.01 & 0.89 $\pm$ 0.01 & $3.35^{+1.37}_{-1.06}$ &\nodata    \\
080823429 & 94     & 14.24   &	PL & 0.09 $\pm$  0.009 & 0.97 $\pm$  0.005 & $4.17^{+1.56}_{-1.28}$ &\nodata  \\
080823478 &   264   &  512.6  & BPL & 0.003 $\pm$  0.002 & 0.13 $\pm$ 0.01 & $2.16^{+2.09}_{-0.84}$ & $5.21^{+2.41}_{-3.25}$   \\
080823623  & 220    & 21.12   & PL & 0.30 $\pm$  0.01 & 0.23 $\pm$  0.01 & $1.97^{+0.55}_{-0.46}$ &\nodata \\
080823714   & 406   &  33.04  & PL & 0.58 $\pm$  0.02 & 0.57 $\pm$  0.02 & $1.77^{+0.34}_{-0.31}$ &\nodata  \\ 
080823847a   & 264 &  78.61  & PL & 0.10 $\pm$  0.009 & 0.63 $\pm$  0.02 & $2.55^{+0.33}_{-0.30}$ &\nodata \\ 
080823847b & 108  &   33.09  & PL & 0.92 $\pm$  0.008 & 0.96 $\pm$  0.005 & $2.48^{+0.55}_{-0.48}$  & \nodata\\
080823986  & 60   &  4.37 & flat & 0.22 $\pm$ 0.01 & \nodata &\nodata &\nodata \\
080824346  & 34  &   5.70 & PL & 0.99 $\pm$  0.003 & 0.78 $\pm$  0.01 & $3.02^{+3.45}_{-1.58}$ & \nodata 	 \\
080824828 &  82  &   6.39  & flat &0.42 $\pm$  0.02 &  0.86 $\pm$  0.01 & \nodata & \nodata \\
080825401  & 128 & 104.8 & PL & 0.14 $\pm$ 0.01 & 0.75 $\pm$ 0.01&  $2.25^{+0.24}_{-0.22}$  & \nodata  \\
080826136  & 160  & 507.3 & BPL & 0.026  $\pm$ 0.005  &  0.99 $\pm$ 0.001 & $2.02^{+0.89}_{-1.41}$ & $4.86^{+2.82}_{-3.00}$  \\
080826236  & 88    & 17.08 & PL & 0.99  $\pm$  0.003 & 0.44  $\pm$  0.02 & $2.27^{+0.69}_{-0.57}$ &\nodata \\
080828875   & 72   &  5.28 & PL & 0.93  $\pm$  0.008 & 0.92  $\pm$  0.008 & $3.42^{+3.09}_{-1.51}$  & \nodata  \\
080903421   &  50 & 10.96 & PL & 0.96  $\pm$  0.006 & 0.76  $\pm$  0.01 & $5.20^{+2.52}_{-2.81}$  &\nodata   \\
080903787   & 100& 13.88 & PL & 0.06  $\pm$  0.007 & 0.66  $\pm$  0.02 & $2.44^{+0.81}_{-0.61}$ &\nodata    \\ 
\enddata
\tablecomments{Burst lengths and fluences are taken from \citet{2011ApJ...739...87L}. The posterior probability for the likelihood ratio is always for the simpler model tested (i.e. either power law or constant). $\alpha_1$ is the power law index in the simple power law, and the low-frequency power-law index in the broken power law case. $\alpha_2$ is the high-frequency power law index in the broken power law case. We quote the fifth and ninety-fifth percentiles for each quantity derived from a MCMC sample of 50000 individual parameter sets.}
\label{tab:bayes_results}
\end{deluxetable*}

\begin{deluxetable*}{|l|r|r|l|r|r|r|r|l|}
\tablewidth{500pt}
\tablecolumns{9}
\tablecaption{RESULTS for the search for periodicities and quasi-periodicities in the entire sample of bursts.}
\tablehead{
\colhead{Burst ID} & \multicolumn{3}{c}{maximum measured power} & \multicolumn{4}{c}{sensitivities [\% rms]} & \colhead{$p(LRT)$} \\
\colhead{} &
\colhead{$T_R$} &
\colhead{$\nu_{\mathrm{max}}$}&
\colhead{$p(T_R)$}&
\colhead{40 Hz} &
\colhead{70 Hz} &
\colhead{100 Hz} &
\colhead{500 Hz} &
\colhead{}}

\startdata
080822529 & 1.43 & 1973 &  0.77 $\pm$ 0.01 &19 & 19 & 19 & 19 &  0.47 $\pm$ 0.02\\
080822647 &  14.36 & 3283 & 0.56 $\pm$ 0.02 &  78 & 40 & 28 & 16	& 0.48 $\pm$ 0.02  \\  
080822981 & 7.79 &  2118 & 0.98 $\pm$ 0.004 & \nodata & \nodata & 72 & 16 &  0.76 $\pm$ 0.02  \\
080823020  &  13.77 & 3145 & 0.31 $\pm$ 0.01 &  80 & 32 &  23 & 13 &  0.50 $\pm$ 0.02  \\
080823091  & 0.44 & 2367 & 0.80 $\pm$  0.01 &  9 & 9 & 9 & 9 &  0.51 $\pm$ 0.02 \\
080823174 & 18.14 & 1711 & 0.24 $\pm$  0.01 & 13 & 12 & 12 & 12 &  0.48 $\pm$ 0.02   \\
080823248 & 12.49 & 95 & 0.91 $\pm$ 0.01 	& 18 & 17 & 17 & 17 &   0.53  $\pm$ 0.02   \\
080823293a & 15.51 & 2069 & 0.32 $\pm$  0.02 &  23 & 14 & 12 & 10 &  0.49 $\pm$ 0.02   \\  
080823293b  &  11.03 & 3593 & 0.54 $\pm$  0.02 & 35 & 35 & 35 & 35	 & 0.50 $\pm$ 0.02 	\\ 
080823319 &14.39 & 1542 & 0.43 $\pm$  0.02 &  28 & 18 & 16 & 14	 & 0.54 $\pm$ 0.02  \\
080823330  & 15.07 & 3695 & 0.46 $\pm$  0.02 & 40 & 19 & 12 & 6 	& 0.88 $\pm$ 0.01\\
080823354  & 12.13 & 2407 & 0.72 $\pm$ 0.01 & 46 & 26 & 21 & 18 & 0.81 $\pm$ 0.02   \\
080823429 &  12.97 & 3689 & 0.55 $\pm$  0.02 & 81 & 24 & 17 & 13 & 0.49 $\pm$ 0.02 \\
080823478  & 15.71 & 2317 & 0.42 $\pm$ 0.02  & 28 & 12 & 8 & 4 &  0.51 $\pm$ 0.02 	\\
080823623  & 18.60 & 902 & 0.09 $\pm$  0.009 & 24 & 17 & 16 & 14 & 0.51 $\pm$ 0.02 	\\
080823714  & 15.03 & 1301 & 0.69 $\pm$  0.01 & 19 & 14 & 13 & 11 & 0.85 $\pm$ 0.01  	\\
080823847a & 18.88 & 4057 & 0.11 $\pm$  0.01 & 43 & 23 & 15 & 8 & 0.49 $\pm$ 0.02  	\\ 
080823847b & 11.03 & 2515 & 0.94 $\pm$  0.007 & \nodata & 57 & 40 & 15	& 0.50 $\pm$ 0.02  \\
080823986   & 10.07 & 2791 & 0.74 $\pm$ 0.01 & 19 & 19 & 19 & 19 & 0.47 $\pm$ 0.02  \\
080824346  & 9.39 & 2968 & 0.83 $\pm$  0.01 & \nodata & 70 & 55 & 26 & 0.52 $\pm$ 0.02    \\
080824828 & 2.54 & 74 & 0.66  $\pm$  0.01 & 	23 & 23 & 23& 23	& 0.24 $\pm$ 0.02  \\
080825401  & 12.36 & 496 & 0.80 $\pm$ 0.01 & 73 & 36 & 26 & 7 & 0.94 $\pm$ 0.007 \\ 
080826136  & 12.80 & 2868 & 0.77 $\pm$  0.01 &  43 & 20 & 11 & 5	 & 0.54 $\pm$ 0.02   \\
080826236  &  13.16 & 3536 & 0.49 $\pm$ 0.02 &  70 & 38 & 28 & 15& 0.56 $\pm$ 0.02 \\
080828875  &  12.24 & 667 & 0.56  $\pm$  0.02 & 54 & 25 & 17 & 17 & 0.53  $\pm$ 0.02  \\
080903421  &  	9.97 & 3781 & 0.66   $\pm$  0.02 & \nodata & 33 & 24 & 22  & 0.56  $\pm$ 0.02\\
080903787  &  14.04 & 2817 & 0.40  $\pm$  0.02 &  73 & 42 & 28 & 16 & 0.50 $\pm$ 0.02 
\enddata
\tablecomments{We show the $T_R = \mathrm{max}_j(\hat{R_j})$ statistics for each burst, along with the associated frequency and the posterior probability to find this outlier given a pure noise process. For each burst, we also quote sensitivities, i.e. the fractional rms amplitude a periodic process would have needed to have in order to be detectable for our method, given the noise process and parameters determined for that burst. Note that due to the excess power in the low-frequency part spectrum being modeled as red noise, the sensitivity will depend on frequency, and be generally less constrained in the low-frequency part of the spectrum than in the white-noise dominated high-frequency spectrum. Where no sensitivity is given, the derived value exceeded 100 \%. A signal with more than 100\% fractional rms amplitude would have negative counts, and is therefore not physical. A sensitivity limit on the amplitude $>100\%$ merely indicates that we cannot constrain the signal amplitude at the given frequency at all. Finally, we also present the posterior probability on the likelihood ratio for a model containing a QPO versus a model without QPO, which is an indicator for the presence of a QPO in the spectrum.}
\label{tab:findper_results}
\end{deluxetable*}

For all bursts, we followed the same procedure as for {\it 080823478}. All of the preferred models had a fairly high $p(SSE)$, which indicates that the overall fit of the preferred model to the data is good. A summary of the results is presented in Table \ref{tab:bayes_results}.
Periodicity searches on the data are summarized in Table \ref{tab:findper_results}. While we compute posterior p-values for all binned and smoothed spectra, we only report the results for the unbinned spectra here for reasons of brevity, and only point out significant results in the binned spectra where appropriate.

\begin{figure*}
\begin{center}
\includegraphics[width=18cm, height=6.5cm]{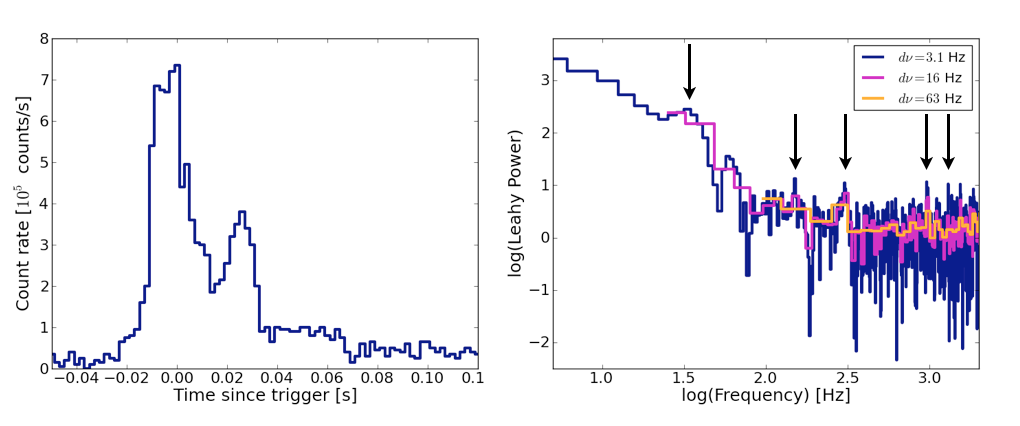}
\caption{Fermi GBM observation of burst {\it bn080823847a} from SGR J$0501+4516$. Left: light curve with a time resolution of 0.002 seconds. Structure in the burst profile is clearly visible. Right: unbinned (blue) and binned (magenta: 16 Hz binning; orange: 65 Hz binning) periodogram for this burst. There is a feature in the periodogram around 30 Hz (leftmost arrow), which is by itself not significant. However, significant features reported in Table \ref{tab:080823847a_detecs} are all at integer multiples of this frequency (within the uncertainty imposed by the frequency resolution), indicating the presence of harmonics at 150 Hz, 300 Hz, 900 Hz and 2100 Hz (arrows 2-5).}
\label{fig:080823847a_lcps}
\end{center}
\end{figure*}

None of the 27 bursts shows periodicities of any noteworthy significance in any of the unbinned  (see Table \ref{tab:findper_results}, column $p(T_R)$) periodograms. The highest data/model outlier significance is seen in burst {\it bn080823847a} (see Figure \ref{fig:080823847a_lcps} for a light curve and periodogram), $p(T_R) = 0.11 \pm 0.01$, at frequency $\nu_{\mathrm{max}} = 4057 \, \mathrm{Hz}$ with a power $P(2I/S) = 18.88$, well below the power required to reach the detection threshold corresponding to a posterior p-value of 5\%. However, the same burst shows significant signals in the binned periodograms, as summarized in Table \ref{tab:080823847a_detecs}. Note that p-values quoted there are corrected for the number of frequencies searched, but neither for the number of bursts searched nor the number of binned spectra searched for each burst. While the former is straightforward (a simple multiplication factor of 27 for the number of bursts searched), the latter is more complicated, owing to the fact that searching different binnings for a single periodogram does not result in independent trials. The most conservative assumption is to consider them independent, including another multiplication factor equal to the number of binnings searched (here: 9). This would rule out all but the two signals with frequency bins of $95$ and $158$ Hz, which remain significant even after a correction for the number of trials.

\begin{deluxetable}{|r|r|r|r|}

\tablewidth{200pt}
\tablecolumns{4}
\tablecaption{Posterior summary for various binned periodograms derived from burst {\it 080823847a}}
\tablehead{\colhead{$d\nu_{\mathrm{bin}}$ [Hz]} &
\colhead{$\nu_{\mathrm{max}}$ [Hz]} &
\colhead{$T_R$ } &
\colhead{$p(T_R)$} }
 \startdata
16	&	310	& 	7.24 	& 	0.017 $\pm$ 0.004 \\
22	&	2094	&	7.01	&	0.004 $\pm$ 0.002 \\
32	&	301	&	6.01	&	0.003 $\pm$ 0.002 \\
47	&	307	&	4.90	&	0.002 $\pm$ 0.001 \\
63	&	4067	&	4.27	&	0.003 $\pm$ 0.002 \\
95	&	2096 &	4.10	&	($<2.0 \times 10^{-5}$) \\
158	&	2129 &	3.26	&	($<2.0 \times 10^{-5}$) \\
316	&	2050	 &	2.59	&	0.027 $\pm$ 0.005 \\
633	&	2050	 &	2.45	&	0.003 $\pm$ 0.002 \\
949	&	2050	 &	2.34	&	0.007 $\pm$ 0.003 \\
1583	&	2050	 &	2.16	&	0.029 $\pm$ 0.005 
\enddata
\tablecomments{P-values were derived using an increased number of $50000$ simulations in order to increase the resolution on small probabilities. The first column holds the binned frequency resolution, $d\nu_{\mathrm{bin}}$, the second column the frequency at which the highest outlier $T_R$ was found, $\nu_{\mathrm{max}}$, column three the corresponding value of the $T_R$ statistic and finally column four the associated posterior p-value to find that value in a pure noise spectrum, binned to the same frequency resolution. Note the probabilities without errors in brackets at binning frequencies of 95 and 158 Hz. The p-value there turned out to be zero at these binning frequencies. Of course, the p-value is not actually zero, however, since we approximate the posterior distribution of $T_R$ with a finite number of simulations, there is a possibility that the true probability to achieve the observed value with only noise is small enough that none of the simulations will exceed the observed $T_R$, giving rise to a zero p-value. We computed the p-values with up to 50000 simulations, and hence state an upper limit on the p-value of $2.0 \times 10^{-5}$.}

\label{tab:080823847a_detecs}
\end{deluxetable}

Comparing the results in Table \ref{tab:080823847a_detecs} to the periodogram of the same burst in Figure \ref{fig:080823847a_lcps} allows for several interesting observations, whose implications will be discussed in detail in Section \ref{sec:discussion}. The frequencies at which significant excess powers are detected in the binned spectra are all at integer multiples of a suspicious feature at around 30 Hz, which in itself is not significant in any of the searches. The periodogram itself shows fairly prominent features at the frequencies at which signals are detected (see arrows in Figure \ref{fig:080823847a_lcps}, right panel), which become more prominent in the binned spectra, lending confidence that these might be real signals, and not false positive detections. If indeed there is a feature at 30 Hz, whose significance is diminished by the presence of red noise, then the higher-frequency detections may correspond to harmonics of this signal. The implications of these findings are discussed in more detail in Section \ref{sec:discussion}. Since we only search for the highest peak in each periodogram, there is a chance that several frequencies may be significant in each binned periodogram. This would require a more extensive search, including e.g. the second- and third-highest peaks in the analysis. Additionally, a potential signal may have an energy dependence, thus an energy-resolved timing analysis may yield more conclusive results. \\ 

In none of the twenty-seven bursts do we find any significant QPOs (see last column in Table \ref{tab:findper_results}). Posterior probabilities for the broadband model alone are largely in the range $0.2$ to $0.8$, indicating that the broadband model alone is an adequate fit to the data, and an additional Lorentzian does not result in a better fit. However, we also note that the posterior probability of the likelihood ratio is clustered around 0.5 for 19 out of 27 bursts. Since for a well-behaved probability statistic applied to data consistent with the null hypothesis, p-values should be uniformly distributed between 0 and 1, this clustering indicates that the test is conservative in the sense that it does not overstate the rejection of either null hypothesis or alternative hypothesis. In practice, results on the likelihood ratio test should be combined with those on the binned spectra to yield reliable detections.

\subsection{Broadband Variability}

The broadband variability observed in the bursts is not just a nuisance when searching for (quasi-) periodicities, but is of interest in its own right: it shows that something is varying in the source, although not periodically. Here, we have chosen a purely phenomenological approach, selecting empirical models that are both simple and widely observed in many astrophysical contexts, without physical justification. Hence, the question of whether we can derive any physical knowledge from these empirical models is an interesting and important one.

 \begin{figure}[h]
\includegraphics[width=9cm]{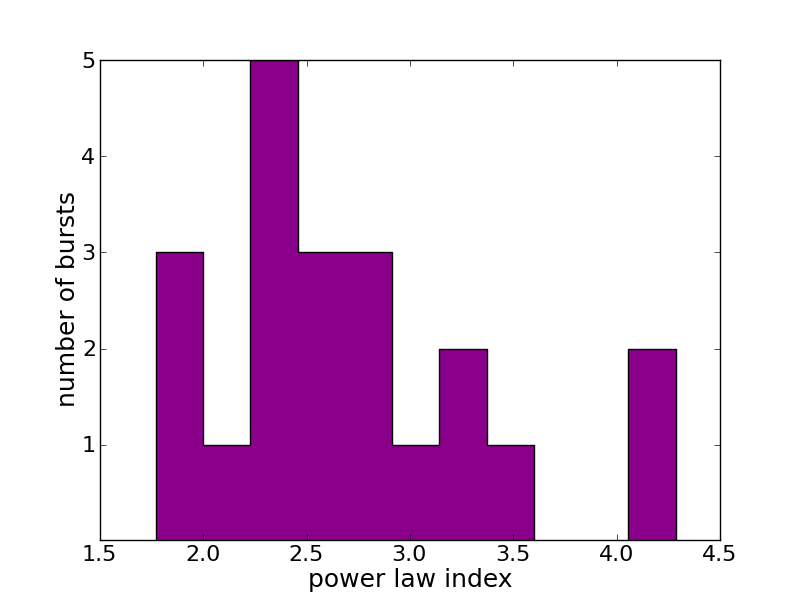}
\caption{Distribution of power law indices ($\alpha$ for single power law, $\alpha_2$ for broken power law) for all bursts where the power law or broken power law was preferred.}
\label{fig:plind_dist}
\end{figure}
Almost all bursts in the sample are well-modeled by a simple power law, although the lower and upper bounds on the 90\% credible interval show a large variation in some indices, indicating that they are not very well constrained. Some caution should be used when interpreting these values, since they were derived from a sample of bursts with very diverse properties overall, such as burst length and fluence. Additionally, an unmodeled burst envelope may significantly change the slope of the power law-like part of the periodogram. A reliable characterisation of the broadband properties hinges on our knowledge of the processes (both noise and non-stochastic) involved, and will be deferred to a future paper involving a larger sample of bursts. 
Figure \ref{fig:plind_dist} shows the distribution of power law indices for all bursts where at least a simple power law was required to adequately represent the data. The distribution of indices ranges from 1.7 to 4.3 and peaks around 2.5, which is higher than commonly seen for example in Gamma-ray bursts (see e.g. \citealt{2000ApJ...535..158B} and \citealt{2012MNRAS.422.1785G}).

Four bursts could be modeled without invoking the presence of red noise at all; contributions by burst variability were confined to very low frequencies and standard Fourier techniques apply for all but the first three or four frequency bins. Only two bursts required the more complex model (bursts {\it 080823478} and {\it 080826236}). While these were not the longest bursts, they had the highest fluence (except for the excluded saturated bursts), indicating a potential correlation between power spectral shape and burst fluence. This is expected: the normalization (i.e. the relative strength to the noise) of the broadband noise model depends directly on the number of counts detected, thus bright bursts may enable us to see the cut-off frequency of the power-law as set by the burst duration, whereas many of the other bursts have too low a fluence to observe the same behaviour. Alternatively, it is possible that the difference in power spectral shape is intrinsic to the source, that is, the physical processes creating this kind of variability vary in some way with burst fluence. Without a model for the emission processes producing the burst in the first place, however, it is difficult to assess the validity of the latter hypothesis. Additionally, with only two bursts preferring the broken power-law model, the numbers are too low to draw conclusions, and we defer the discussion on the physical implications of the broadband noise modeling to a later paper utilizing a larger sample of magnetar bursts. 

\section{Discussion and Conclusion}
\label{sec:discussion}

Magnetar bursts are a potential window into the interior of neutron stars, via the oscillations measured in magnetar giant flares. Finding analogous signals in the wealth of short SGR bursts, however, poses something of a challenge.
We have shown that timing analysis of astrophysical transients is a non-trivial problem. Standard Fourier techniques are defined for infinitely long time series, an assumption that is clearly broken by the non-stationary nature of transient events in general, and magnetar bursts in particular.  

Monte Carlo simulations of light curves fail to be predictive when there is no precise knowledge of the underlying burst light curve: there is a degeneracy between the overall, aperiodic burst shape, a potential red noise component, and the very thing we would like to measure: a QPO. When the light curve is not adequately modeled, then the periodograms produced from the Monte Carlo simulations will not reproduce the low-frequency part of the periodogram, where it clearly diverges from the statistical distributions expected for pure white noise. 

In the absence of better knowledge about the emission processes in magnetar bursts, we advocate a conservative Bayesian method that models the burst light curve as a pure red noise process. It is purely empirical in the sense that it does not require additional assumptions on the underlying physical processes, except for fairly broad assumptions on what the power spectrum shape might be. 
Assuming pure red noise is, in a way, the complementary extreme to Monte Carlo simulations of the light curve: in one, we assume only a deterministic burst profile without the presence of red noise, with the price that inadequate modeling of the periodogram shape will lead to spurious detections. Here, we assume only red noise, at the cost that weak signals are likely missed. This is the greatest weakness of our approach. We have shown in Section \ref{sec:burstenvelope} that even strong signals may be undetectable at low frequencies, where burst envelope and red noise dominate. These, however, are exactly the frequencies at which many of the QPOs in giant flares have been seen (e.g. 18 Hz, 30 Hz for the 2004 giant flare, see \citet{2005ApJ...628L..53I}). This limitation is in part not only due to restrictions of our method, but also to the short lengths of the SGR bursts, where at these frequencies only one or two cycles may be seen in the light curve. Upper limits below 100 Hz are often fairly unconstraining, and range from 10 \% to more than 100 \% fractional rms amplitude for a signal to be detectable. At frequencies close to and above 100 Hz, sensitivities approach the white noise limit, which is strongly dependent on the number of photons from a particular burst. Thus, for a bright burst with good count statistics, sensitivities are quite constraining, down to less than 10 \% (e.g. burst {\it 080823478}, see Table \ref{tab:findper_results}). This is comparable to what was observed in giant flares: for example, a QPO at 93 Hz, as seen in the 2004 flare, at roughly 10 \% rms amplitude \citep{2005ApJ...628L..53I, 2006ApJ...637L.117W}, should be detectable in at least the brightest bursts of our sample. Similarly, a high-frequency QPO like the one at 625 Hz seen in the 2004 flare with a fractional rms amplitude of up to 20 \% should be clearly detected with our method as well.

However, QPOs in SGR bursts may be less strong than in the giant flares, owing to the lower energy injected in SGR bursts, and hence more likely to be misclassified as non-detections, if their fractional rms amplitudes fall below 5 \%. Additionally, we restrict ourselves when searching for periodicities by considering only the highest peak in the spectrum, which is clearly not adequate when there are multiple signals in the periodogram. On the other hand, if even the highest peak is not significant, any other peak in the periodogram will be even less significant. 

The burst {\it 080823847a} presents an interesting case that illustrates the limits of a pure signal-processing approach to the timing analysis shown here. Two of several significant signals detected in the binned spectra remain significant even after the most conservative correction for the number of trials is applied, indicating that there is indeed a real signal present. However, the nature of this signal is at present unclear. The detected signals are possibly harmonics of a lower-frequency signal around $30$ Hz, corresponding to a timescale of $\tau = 1/\nu = 33$ ms. This timescale roughly corresponds to the two sharp peaks seen in the burst light curve in Figure \ref{fig:080823847a_lcps} (left side). Whether we consider this to be a QPO atop a burst envelope or not cannot be answered from Fourier analysis alone; it becomes a matter of interpretation and prior knowledge. The presence of harmonics indicates a signal repeating on the time scale of the fundamental, but with variability on shorter timescales in the signal. One could well interpret the two peaks in the light curve as a strongly damped (quasi-)periodic signal that is amplified together with some underlying burst profile and dies away after two, possibly three, cycles. The frequency of this signal is similar to that observed from the 2004 giant flare (\citealt{2005ApJ...628L..53I}, \citealt{2005ApJ...632L.111S}), and thus not unlikely. On the other hand, this kind of signal can equally well be derived from a red noise process. The fact that red noise is a stochastic process means that at some point, two or even three peaks will follow each other, as in the present burst. While red noise itself would not introduce harmonics, the signal could be boosted by an underlying burst envelope, introducing the observed harmonics. At present, without any knowledge about emission processes and the kind of light curve they produce, it is impossible to distinguish the two, and we choose the conservative approach and interpret the observed feature as part of a noise process. 

Another problem as yet unsolved is that of false non-detections we expect, i.e. weak signals missed due to the fact that we assume a pure red noise process. There are several ways to break this dilemma, but all require more detailed knowledge of the variability-producing processes in the neutron star, and this is where both theoretical efforts and development of novel statistical techniques are required. What produces the burst emission? What produces the aperiodic variability seen in the red noise part of the periodograms? Until we can answer these questions, finding QPOs in magnetar bursts will always suffer from the essential degeneracy between burst envelope and red noise. If we knew the overall burst shape, one could for example simulate light curves, but as a combination of a burst profile and a red noise process, as done in Section \ref{sec:burstenvelope}, and compare this sample to the observed periodogram. Other approaches involve leaving behind the Fourier domain and its incorrect assumption of stationarity behind altogether. 

Knowledge about the burst envelope, on the other hand, would also offer us an additional source of observational information to exploit: if we can use the existing information on the hundreds of bursts available to learn something about the burst envelope shape, we may be able to put tight constraints on potential QPO detections and provide additional observational constraints for burst emission models in general. 
Clearly, with the right statistical techniques, there is a wealth of information yet to be extracted from the SGR bursts observed with Fermi GBM. Additionally, for bursts with high count rates, it is possible to study variability properties of the bursts with energy, thanks to Fermi/GBM's excellent energy resolution. These studies may provide additional information on QPOs that depend on energy. 
The methods developed here, however, while developed with SGR bursts in mind, are by no means limited to magnetars. They are applicable in fairly general circumstances, for any light curve that is phenomenologically similar to what we observe from magnetars: highly variable, transient events with complex light curves. This includes, for example, other known transients such as gamma-ray bursts (GRBs), tidal disruption events and supernova light curves. \\

\acknowledgments
D.H. and A.L.W acknowledge support from a Netherlands Organization for Scientific Research (NWO) Vidi Fellowship (PI A. Watts), and would like to thank Jason Farquhar for useful discussions. C.K. and was partially supported by NASA grant NNH07ZDA001-GLAST.  E.G. acknowledges support from the Scientific and Technological Research Council of Turkey (T\"UB\"ITAK) through grant 109T755.  

\bibliography{sgr0501_paper_references}
\bibliographystyle{apj}

\appendix
\label{sec:appendix}

\section{Comparison with Recent Results}

\begin{figure*}[h!]
\begin{center}
\includegraphics[width=17cm]{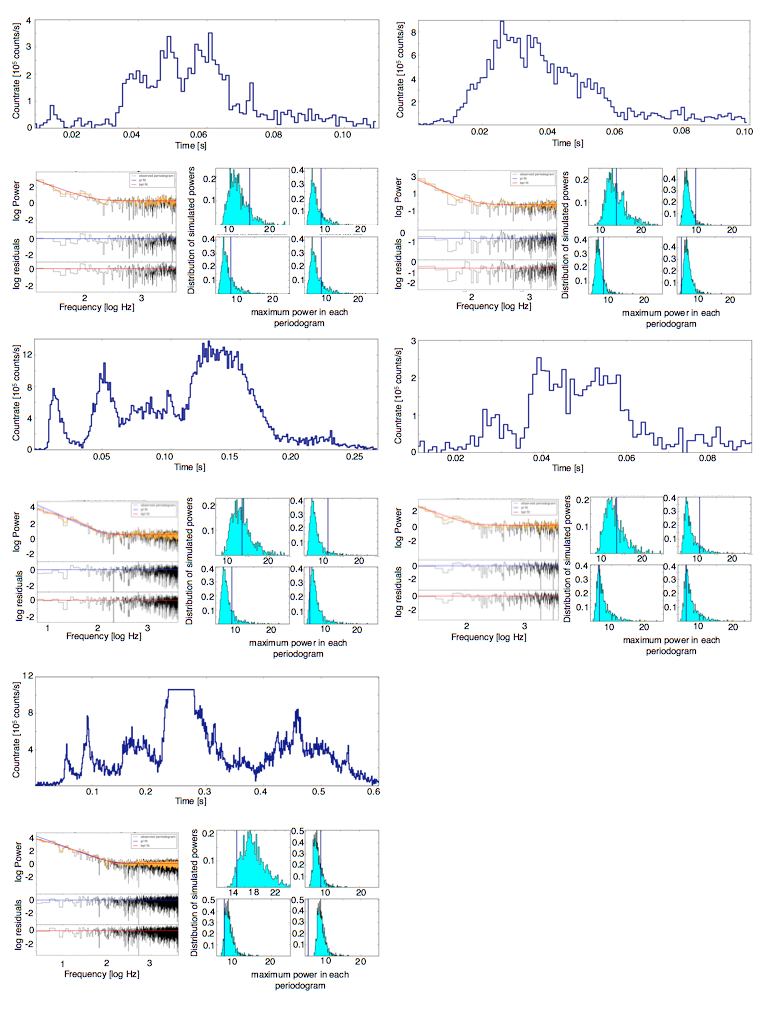}
\caption{Light curves, periodograms and posterior distributions for the five bursts from SGR J$1806-20$ observed with the Rossi X-ray Timing Explorer between 1996 November 5-18 presented in \citet{2010ApJ...721L.121E}. Long, upper plots for each burst are the light curves binned to $0.005$ seconds. On the lower left, the unsmoothed periodograms (black) the five-bin smoothed periodogram (orange) as well as the power law and broken power law fits. Underneath each periodogram is a plot of the residuals of dividing the periodogram by the broadband model (power law in the middle, and bent-power law on bottom). On the right, the posterior distributions of the highest data/model outlier for the unsmoothed (upper left) and smoothed periodograms (rest; upper right: 3-bin smoothing, lower left: 5-bin smoothing; lower right: 11-bin smoothing). The observed value is overplotted as a black vertical line.}
\label{fig:emresults}
\end{center}
\end{figure*}

\citet{2010ApJ...721L.121E} searched a data set of SGR bursts from SGR J1806-20, found in data taken with the Rossi X-ray Timing Explorer (RXTE). They report the significant detection of QPOs in five different SGR bursts out of a sample of thirty, with frequencies of $84$, 103 and 648 Hz with significances $\geqslant 4.3 \, \sigma$, estimated using Monte Carlo simulations of light curves equivalent to those described in Section \ref{sec:mc}, by smoothing the burst light curve and subsequently adding Poisson detector noise to form a null hypothesis against which to test the data. For the reasons stated in Section \ref{sec:mc}, we do not believe these estimates to be conservative, and consequently reanalyze this data set from 1996 November 5-18 after barycentering the data and filtering out photons outside the range $2-60 \mathrm{keV}$, where the response curve of the instrument indicates that noise will dominate at these energies. We use data from all proportional counter units (PCUs) of the PCA detector. 
We use the Bayesian analysis presented in Section \ref{sec:bayes} to choose a broadband noise model and search for (quasi-)periodic signals in the same data set. 
The results are presented in Table \ref{tab:emresults} and Figures \ref{fig:emresults} (a) to (e).

%
%
\begin{table*}[h!]
\caption{Summary of the Bayesian analysis for five bursts presented in \citet{2010ApJ...721L.121E}. The choice of broadband model is recorded, as well as the results of the periodicity and QPO searches. Note that the posterior probability for the summed squared residuals ($p(SSE)$ is shown for the model with the better fit, whereas the posterior probability for the likelihood ratio $p(LRT$) is shown for the simpler model always.}
\begin{center} 
\begin{tabular}{|c|c|c|c|c|c|c|} \hline
&quantity & burst 1 & burst 2 & burst 3 & burst 4 & burst 5 \\ \hline
&start time [MET s] & 90915519.65 & 90909708.72 & 90925017.31 & 90915519.65 & 9093707656 \\
&length [s] & 0.13 & 0.31 & 0.12 & 0.11 & 0.36 \\\hline
&model & PL & PL & PL &  PL & PL  \\
&p(LRT) & 0.84 $\pm$ 0.01& 0.59 $\pm$ 0.01 &  0.13 $\pm$ 0.01 &  0.41 $\pm$ 0.01 &  0.38 $\pm$ 0.02\\
& p(SSE) & 0.91 $\pm$ 0.008 &  0.85 $\pm$ 0.01 & 0.39 $\pm$ 0.02 & 0.26 $\pm$ 0.01& 0.75 $\pm$ 0.01\\\hline 
& $\alpha_1$  & 2.63 & 2.68 &  2.52 &  2.67 & 2.29\\
& $\alpha_1$ 5\%  & 2.07 & 2.31 &  2.06 & 2.10 & 2.11 \\ 
& $\alpha_1$ 95\%  & 3.31 & 3.12 &  3.04 & 3.37 &  2.47  \\\hline
\multirow{3}{*}{unsmoothed}& max(2I/S) & 15.11 & 13.99 & 13.67 & 13.18 & 14.46  \\
& $\nu_{\mathrm{max}}$ & 2464 & 2690 &  3696 &  3694 & 3246 \\
& $p(T_R)$ &  0.13 $\pm$ 0.01 & 0.41 $\pm$ 0.02 & 0.08 $\pm$ 0.01  & 0.32 $\pm$ 0.01 & 0.95 $\pm$ 0.02 \\\hline

\multirow{3}{*}{3 bins}&  max(2I/S) & 8.34 & 9.21 &  8.65 & 9.4 & 9.24  \\
 & $\nu_{\mathrm{max}}$ &  2466 & 2701 &  3706 & 3698 & 1394  \\
 & $p(T_R)$ &  0.20 $\pm$ 0.01 & 0.05 $\pm$ 0.01 & 0.03 $\pm$ 0.01 &  0.11 $\pm$ 0.01 &  0.18 $\pm$ 0.01\\\hline

\multirow{3}{*}{5 bins} & max(2I/S) & 7.38 & 8.51& 8.73 & 6.33 & 7.15  \\
 & $\nu_{\mathrm{max}}$ & 2352 &  2703 & 3695 & 3687 &  1395 \\
& $ p(T_R)$ & 0.11 $\pm$ 0.01  & 0.05 $\pm$ 0.01 & 0.11 $\pm$ 0.01 & 0.24 $\pm$ 0.01& 0.20 $\pm$ 0.01 \\\hline

\multirow{3}{*}{11 bins} & max(2I/S) &  4.27 & 6.91 & 5.05 & 4.98 & 5.85  \\
&  $\nu_{\mathrm{max}}$ & 2416 & 2708 & 3712 &  3648 & 2218 \\
 &$p(T_R)$ & 0.07 $\pm$ 0.01 & 0.41 $\pm$ 0.01 &  0.23 $\pm$ 0.01 & 0.29 $\pm$ 0.02 & 0.53 $\pm$ 0.01  \\\hline
 
 & QPO $p(LRT)$ &  0.49 $\pm$ 0.02 & 0.20 $\pm$ 0.02 & 0.47 $\pm$ 0.02 & 0.54 $\pm$ 0.02& 0.25 $\pm$ 0.01 \\ \hline
\end{tabular}
\end{center}
\label{tab:emresults}
\end{table*}

For the most part, we cannot confirm the detections shown in \citet{2010ApJ...721L.121E} using our Bayesian methods.  Only one burst shows a marginally significant signal: burst 3 ($p = 0.03 \pm 0.01$) at a frequency of around 3706 Hz, and only in one binned spectrum. Given the posterior probability is very close to our (rather high) detection threshold, we are inclined to disregard this detection as insignificant as well, as it would indeed become insignificant as soon as we take the number of bursts searched into account. This result is in stark contrast with the probabilities quoted in \citet{2010ApJ...721L.121E}. There are, however,  errors in their analysis: taking into account the varying nature of the background lightcurve should make the significance of any claimed detection drop compared to the significance computed from the ideal $\chi^2$ distribution. Although \citet{2010ApJ...721L.121E} do carry out simulations, the significances that they quote rise substantially, indicating a problem in their Monte Carlo simulation method. Indeed their simulated power spectra show far fewer high noise powers than one would expect given the number of simulations carried out and the number of independent frequency bins (enhancing the significance of any tentative detection). 

\section{Data Analysis Recipes}
\label{sec:recipes}

\subsection{How to fit a noise model}
\label{sec:noise_recipe}
Fitting a noise model is essentially a model selection task. In the following, we will lay out the individual steps in a recipe-like style.

\begin{enumerate}
\item Compute a periodogram of the burst light curve.
\item Fit the periodogram with both a simple model (the null hypothesis) and a more complex model (the alternative hypothesis we wish to test against), to get MAP estimates for the parameters in each model
\item Using the MAP estimates, compute the likelihoods of the data given each model and MAP parameters, then compute the likelihood ratio of the complex model versus the simple (null) model.
\item Produce a large MCMC sample approximating the posterior distributions of the parameters of the null model.
\item Pick a (large) number $n$ of parameter vectors from this sample (e.g. $n= 1000$), and create power spectra from these parameters and simulate a periodogram from each by drawing a realization from the random process the power spectrum represents. This will yield $n$ fake periodograms.
\item Fit each simulated periodogram with both simple and complex model, exactly in the same way as done for the burst periodogram, and compute the likelihood ratio of this simulation. The sample of likelihood ratios from fitting the simulated periodograms will be representative of the likelihood ratios one expects when fitting both the simple and complex model to a sample of periodograms derived entirely from the null hypothesis.
\item Compare the distribution of simulated likelihood ratios with that of the observed burst, and compute the tail area probability of seeing the observed likelihood ratio, if the data were entirely drawn from the null hypothesis. If this tail area probability is large, then the data are consistent with the null hypothesis. The converse, however, is not necessarily true. A small p-value indicates that the data are unlikely to be drawn from the null hypothesis. This is not a direct proof that the complex model is the underlying process that produced the observed burst, however, it is an indication that the more complex model is likely a better representation of the data than the null hypothesis.
\end{enumerate}

\subsection{Searching for periodicities}

\begin{enumerate}
\item Fit a broadband noise model (e.g. the preferred model as defined as above in Section \ref{sec:noise_recipe}) to the burst, and compute the residuals $R_j = 2I_j/S_j$, where $I_j$ are the periodogram powers and $S_j$ is the broadband noise model at $j$th frequency $\nu_j$. 
\item From the residuals, pick the highest outlier  $\max{R_j}$; this is the candidate single-bin periodicity. 
\item Simulate a large number of periodograms from an MCMC sample in the same way as done for the choice of noise model above.
\item Fit each simulated periodogram with the preferred noise model, compute the data/model residuals $R_j$ and find the maximum outlier $\max{R_j}$ in the residuals of each simulated periodogram.
\item Compare the distribution of maximum outliers from the set of simulations derived from the broadband noise model with no periodicity present, to the outlier in the real burst periodogram. One may compute the posterior predictive p-value for the observed outlier in much the same way as for the LRT. If the p-value is large, then the outlier is consistent with a pure noise distribution.
\end{enumerate}

\subsection{Searching for QPOs}

We search for QPOs as a model selection problem, where we compare the broadband noise model to a more complex model combining both the broadband noise model and a Lorentzian to account for the QPO. Because we do not know the centroid frequency of the potential QPO inherently, the task is slightly more complex than that which we use for the choice of noise model. 

\begin{enumerate}
\item Fit the observed periodogram with the broadband noise model and compute residuals $R_j$. Smooth the residuals with a Wiener filter with a width of $5$ frequency bins in order to reduce the probability of the minimisation algorithm terminating in local minima due to sharp noise features.
\item At each frequency, fit a flat line and a Lorentzian of variable width and intensity, but fixed centroid frequency to the smoothed residuals and compute the likelihood of that fit. We leave out the first five and last five bins in order to avoid fitting the edge of the periodogram. The result of this process will be the maximum likelihood as a function of frequency.
\item Pick the frequency bin with the largest maximum likelihood as given from modeling each frequency.
\item Fit the full-resolution periodogram with the broadband noise model alone as well as a combined model of broadband red noise and Lorentzian, using the model parameters for the largest maximum likelihood fit in the previous step as starting parameters for this fit.
\item Compute the likelihood ratio for these two models.
\item Simulate a large number of periodograms (in our case, 500 to reduce computational load) from an MCMC sample of the broadband noise model. 
\item For each simulated periodogram, follow the exact same procedure in steps 1 through 4 to produce an approximation to the distribution of likelihood ratios from the null hypothesis.
\item Compare the distribution of likelihood ratios derived from the simulated periodograms to the likelihood ratio for the observed burst, and compute a posterior predictive p-value for the probability of obtaining the observed likelihood ratio, if the data were consistent with pure red noise.

\end{enumerate}

\end{document}